\documentclass[aps,prd,twocolumn,nofootinbib,groupedaddress]{revtex4-2}

\usepackage{booktabs}
\usepackage{amsmath}
\usepackage{caption} 
\usepackage{array}
\usepackage{siunitx}
\usepackage{amssymb}
\usepackage{graphicx}
\usepackage{caption}
\usepackage{lipsum}
\usepackage{subcaption}  
\usepackage{hyperref}
\usepackage{mathtools}
\usepackage{placeins}

\newcommand{\SOUTHCUT}{\vspace{0.5em} School of Physics and Optoelectronics, South China University of Technology, \\ Guangzhou 510641, People's Republic of China}

\begin{document}

\title{Quasinormal modes and grey-body factors of axial gravitational perturbations of regular black holes in asymptotically safe gravity}

\author{Qi-Long Shi}
\author{ Rui Wang}
\author{Wei Xiong}
\author{Peng-Cheng Li}
\email{pchli2021@scut.edu.cn, corresponding author}

\affiliation{\SOUTHCUT}

\date{\today}

\begin{abstract}

In this paper, we present a detailed study of axial gravitational perturbations of the regular black hole solution in asymptotically safe gravity, as proposed in \cite{Bonanno:2023rzk}. We analyze the quasinormal mode (QNM) spectrum of this black hole using two numerical techniques: the Bernstein spectral method and the asymptotic iteration method (AIM). These approaches allow us to compute QNM frequencies with high accuracy, even for higher overtones. Our results show that the fundamental mode is only weakly affected by the deviation parameter, whereas notable deviations from the Schwarzschild case emerge for higher overtones. Additionally, we examine the correspondence between grey-body factors and QNMs by comparing the WKB grey-body factors with accurate QNM frequencies, finding excellent agreement, especially for larger multipole numbers $\ell$.

\end{abstract}

\keywords{QNMs,regular black hole}

\maketitle

\section{Introduction}

General relativity (GR), despite its profound success in describing gravitation and the large-scale structure of the universe, is not a complete theory. It faces several unresolved issues, including the nature of dark matter and dark energy, the quantization of gravity, and the presence of singularities in spacetime. Among these, the problem of singularities has received particular attention over the decades \cite{Penrose:1964wq,Hawking:1970zqf}. To address this, various approaches have been proposed. A pioneering effort was made by Bardeen, who introduced the first model of a regular black hole (BH) by replacing the mass term in the Schwarzschild solution with a radially dependent function \cite{Bardeen:1968bd}.
Subsequent work extended this idea by constructing regular BH metrics through either the coupling of Einstein's equations with specific matter sources  or by interpreting them as quantum-corrected versions of classical singular solutions (see \cite{Bambi:2023try} for an extended review).

In recent years, regular BH models \cite{Platania:2023srt,Spina:2025wxb} have attracted increasing attention within the framework of asymptotically safe gravity \cite{Bonanno:2000ep} as a promising approach to resolving singularities. A notable development is the work by Bonanno and collaborators \cite{Bonanno:2023rzk}, who, building on earlier ideas by Markov and Mukhanov \cite{Markov:1985py}, derived an explicit metric describing the exterior of a collapsing dust sphere. This result was obtained using an effective Lagrangian with a coupling between gravity and matter, consistent with the asymptotic safety scenario. Unlike traditional approaches that require exotic matter to avoid singularities, their model predicts the formation of regular BHs purely through gravitational dynamics during collapse.

Despite these advances, observationally distinguishing regular BHs from classical singular ones remains highly challenging. In this context, QNMs offer a powerful diagnostic tool. The concept was first introduced by Vishveshwara in the study of BH perturbations \cite{Vishveshwara:1970zz}, where the real and imaginary components of QNM frequencies represent oscillation and damping rates, respectively. These damped oscillations, which characterize the ringdown phase of gravitational wave signals, encode information about spacetime geometry near the event horizon \cite{Nollert:1999ji, Kokkotas:1999bd,Berti:2009kk,Konoplya:2011qq,Bolokhov:2025rng}. Therefore, the measurement of QNMs can be used to test GR and constrain modified gravity theories \cite{Berti:2018vdi,Berti:2025hly}. However, most studies focus primarily on fundamental QNM modes, often neglecting high overtone modes. Recent research has shown that even small modifications near the event horizon can significantly alter the first few overtones \cite{Konoplya:2022pbc}, a phenomenon referred to as the ``outburst'' of overtones. Separately, including higher overtones may enhance the accuracy of ringdown signal modeling \cite{Giesler:2019uxc, Giesler:2024hcr}; see, however, \cite{Baibhav:2023clw} for a different viewpoint. Consequently, increasing attention is being given to these high-overtone modes in various BH models, see e.g. \cite{Konoplya:2022hll,Konoplya:2022iyn,Fu:2023drp,Moreira:2023cxy,Konoplya:2023aph, Konoplya:2023ppx, Konoplya:2023ahd,Bolokhov:2023bwm,Gong:2023ghh,Cao:2024oud,Zhang:2024nny,Gingrich:2024tuf,Hirano:2024fgp,Dubinsky:2024gwo,Bolokhov:2023ruj,Livine:2024bvo,Zhu:2024wic,Tang:2024txx, Lutfuoglu:2025ljm}.

For the regular BH solution introduced by Bonanno et al., existing studies have explored aspects such as BH shadows and QNMs under scalar and electromagnetic perturbations \cite{Stashko:2024wuq, Spina:2024npx, Sanchez:2024sdm}. Although gravitational perturbations have been studied to some extent \cite{Lutfuoglu:2025ohb} \footnote{The gravitational perturbations  of the improved Bonanno-Malafarina-Panassiti metric (which includes an additional term) were studied in \cite{Bonanno:2025dry}.}, accurate determinations of QNMs---especially for high overtone modes---have not yet been explored using high-precision methods, to the best of our knowledge. This paper addresses this gap by investigating axial gravitational perturbations and extracting both fundamental and high overtone QNMs.
On the other hand, QNMs and grey-body factors arise from distinct boundary conditions and describe different physical aspects, a rigorous correspondence between them has been identified, particularly in the high-frequency (eikonal) regime \cite{Konoplya:2024lir, Konoplya:2024vuj}. This connection was motivated by recent findings suggesting that grey-body factors may be related to the amplitudes of BH ringdown signals \cite{Oshita:2023cjz, Okabayashi:2024qbz}, and by the observation that grey-body factors are more robust against small deformations of the effective potential compared to higher overtone QNMs \cite{Rosato:2024arw, Oshita:2024fzf,Konoplya:2025ixm}.
To explore this relationship in broader contexts, such as those done in \cite{Skvortsova:2024msa,Dubinsky:2024vbn,Bolokhov:2024otn,Malik:2024wvs,Malik:2024cgb,Lutfuoglu:2025hjy,Hamil:2025cms,Tang:2025mkk,Pedrotti:2025upg,Liang:2025jph,Xie:2025jbr},  we apply the 6th-order WKB approximation \cite{Konoplya:2003ii} to calculate the grey-body factors. We then compare these results with those obtained by substituting the accurate QNM frequencies into the QNM/grey-body factor correspondence, in order to verify this correspondence for axial gravitational perturbations of the regular BH \cite{Bonanno:2023rzk} with multipole numbers $\ell = 2$ and $\ell = 3$.

The paper is organized as follows. In Sec.\ref{The basic equations}, we present a brief review of the BH spacetime \cite{Bonanno:2023rzk}. In Sec.\ref{Perturbation equation for the gravitational field}, we discuss the wave equations governing axial gravitational perturbations. In Sec.\ref{methods}, we provide a detailed exposition of two numerical methods used in this work for computing QNMs. In Sec.\ref{QUASINORMAL MODES}, we explore QNMs, including fundamental modes and high overtone modes. In Sec.\ref{GREY-BODY FACTORS OBTAINED VIA THE CORRESPONDENCE WITH QUASINORMAL MODES}, we introduce the calculation of greybody factors using the QNMs/grey-body correspondence \cite{Konoplya:2024lir} and test the accuracy of this correspondence with the results obtained form 6th-order WKB approximation. We finally conclude in Sec.\ref{CONCLUSIONS}. The geometric units $ G=1=c$ are maintained throughout this paper. 

\section{Background}
\label{The basic equations}
In this section, for completeness we briefly review the derivation of the regular BHs presented in \cite{Bonanno:2023rzk}.   Bonanno et al. \cite{Bonanno:2023rzk} extended the idea of Markov and Mukhanov \cite{Markov:1985py} and studied the gravitational collapse of dust (pressureless ideal fluid) within the framework of asymptotic safety. By introducing a multiplicative coupling between matter and gravity, guided by the Reuter fixed point \cite{Bonanno:2000ep}, the authors derived an effective Lagrangian in which the gravitational coupling weakens at high energies. This leads to a regular interior solution that smoothly matches a static, asymptotically flat exterior geometry. The starting point is the following action
\begin{equation}
 S=\frac{1}{16\pi} \int d^{4} x \sqrt{-g}[R+2 \chi(\epsilon) \mathcal{L}],
\end{equation}
where $\mathcal{L}=-\epsilon$ is the matter Lagrangian, $\epsilon$ is the energy density of the fluid and $\chi (\epsilon )$ represents the coupling between gravity and matter  with the property $\chi (\epsilon =0)=8\pi$.

As a result, the equations of motion are given by
\begin{equation}\label{EoMcollapse}
	R_{\mu\nu}-\frac{1}{2} R g_{\mu\nu}=8\pi G(\epsilon) T_{\mu\nu}-\Lambda(\epsilon)g_{\mu\nu},
\end{equation} 
where $T_{\mu\nu}$ is the stress-energy tensor of dust, $G(\epsilon)$ and $\Lambda(\epsilon)$ represent the effective Newton constant and cosmological constant, respectively. The effective Newton constant $G(\epsilon)$ is determined  from asymptotically safe gravity, under the key assumption that the Reuter fixed point remains minimally affected by the presence of matter. Once $G(\epsilon)$ is determined, both the coupling function $\chi(\epsilon)$ and the effective cosmological constant $\Lambda(\epsilon)$ can be fixed as well. Then the equations of motion (\ref{EoMcollapse}) can be solved to obtain a regular interior solution by considering the spherically homogeneous collapse of dust.

\begin{figure}[h]
    \centering
    \includegraphics[width=0.45\textwidth]{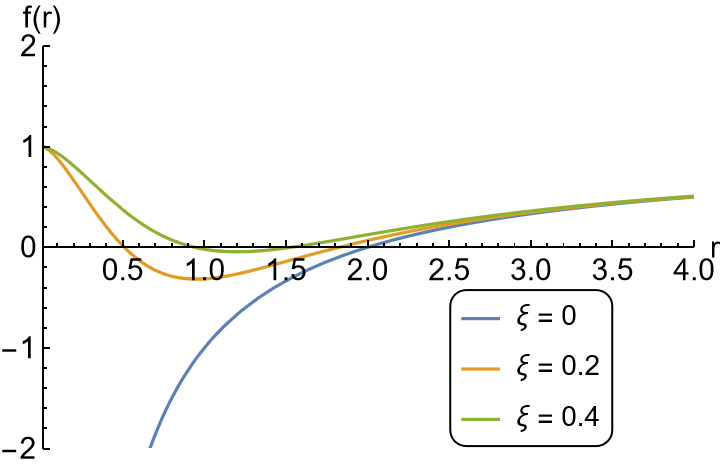} 
    \caption{The metric function $f(r)$ for various $\xi$.}
    \label{fr}
\end{figure}

The static and spherically symmetric exterior geometry of the BH in this situation  is uniquely determined by the condition that both the induced metric and the extrinsic curvature at the collapsing boundary are continuous across the interface, when approached from either side. The metric outside the BH is given by \cite{Bonanno:2023rzk}
\begin{equation}
    \label{bh solution}
 d s^{2}=-f(r) d t^{2}+f(r)^{-1} d r^{2}+r^{2}\left(d \theta^{2}+\sin ^{2} \theta d \phi^{2}\right),
\end{equation}
with 
\begin{equation}
f(r)=1-\frac{r^2}{3\xi} \log\left(1 + \frac{6M\xi }{r^3} \right),
\end{equation}
where $M$ is the BH mass and $\xi$ is a scale parameter characterizing the deviation from the Schwarzschild BH whose value cannot be determined from  first principles and should be constrained from observations.\footnote{Note that the $\xi$ carries dimensions of length squared. If $\xi$ is associated with the Planck scale, its effects are expected to be strongly suppressed for astrophysical BHs. This limitation is common to a wide class of regular BH models motivated by quantum gravity and does not affect the internal consistency of the effective spacetime description. In the present work, $\xi$ is therefore treated as a phenomenological parameter, and our analysis focuses on the structural properties of gravitational perturbations and QNMs rather than on direct observational constraints.} When $\xi\to 0$, the Schwarzschild BH is recovered. By numerically solving the equations $f(r_h)=f'(r_h)=0$, we can obtain the critical value  $\xi _{cr} \simeq 0.4565M^{2}$ and $r_h=1.2516M$. In subsequent calculations, we set $M$ to 1 such that all quantities are dimensionless. The metric function $f(r)$ with various values of $\xi$ is shown in Fig.\ref{fr}. We can see that when $0<\xi<\xi _{cr}$, the  metric has two event horizons, the inner and outer ones. Moreover, the two horizons merge into a single horizon, corresponding to an extreme BH, as $\xi =\xi _{cr}$. When $\xi > \xi_{cr}$, no horizon exists. 

\section{Perturbation equation for the gravitational field}
\label{Perturbation equation for the gravitational field}

To study the QNMs of gravitational perturbations for the regular BH described by (\ref{bh solution}), we need to perturb the underlying equations of motion. However, since the metric (\ref{bh solution}) is obtained by matching the exterior geometry with an exact interior solution, rather than being derived from a known action or as an exact solution of a specific field equation such as (\ref{EoMcollapse}), the fundamental underlying theory remains unknown.	To address this issue, we follow the approach adopted in several works \cite{Toshmatov:2017kmw, Ashtekar:2018cay, Chen:2019iuo, Yang:2023gas, Konoplya:2024lch}, where quantum corrections are effectively modeled as arising from the stress-energy tensor of an anisotropic fluid within the framework of classical Einstein gravity. As we will show below, in this framework the axial (odd-parity)  perturbations are particularly simple, since they decouple from the matter perturbations. In contrast, the polar (even-parity) perturbations are sensitive to the details of the matter sector, which may reflect subtle differences between genuine quantum corrections and their effective fluid description. For this reason, in the present work we restrict our attention to the axial sector. The similar approach was employed to investigate the axial gravitational perturbations of quantum-corrected BHs in various models \cite{Chen:2019iuo, Yang:2023gas, Konoplya:2024lch}.



\begin{figure*}[ht]  
    \centering
    \begin{minipage}[t]{0.48\textwidth}
        \centering
        \includegraphics[width=\linewidth]{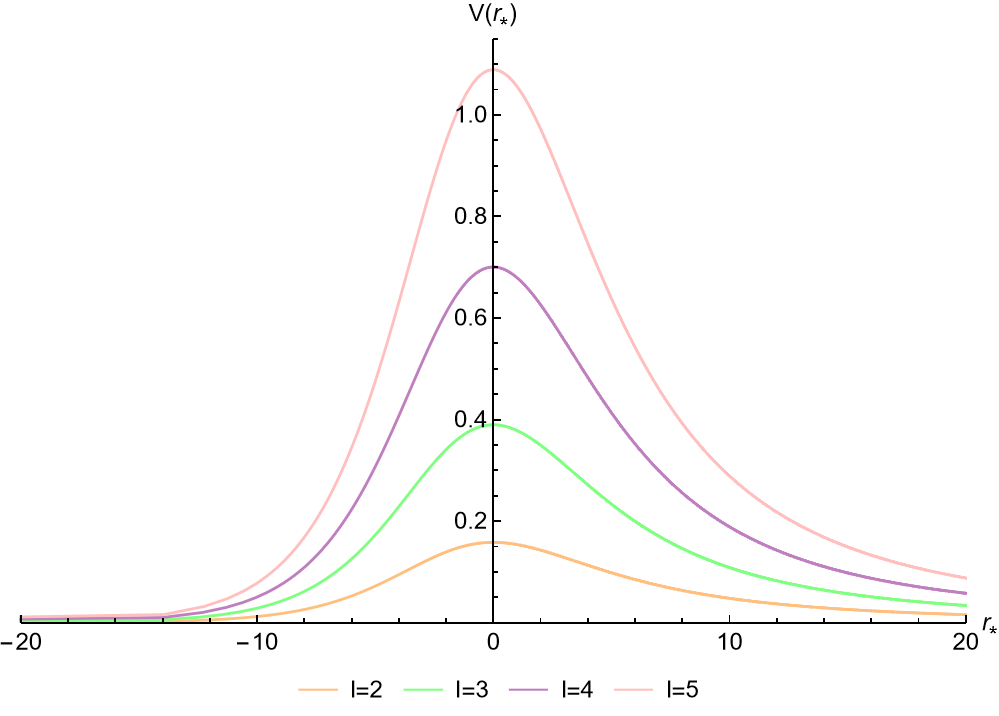}
        \label{fig:xi=0.2vrstar}
    \end{minipage}
    \hfill  
    \begin{minipage}[t]{0.48\textwidth}
        \centering
        \includegraphics[width=\linewidth]{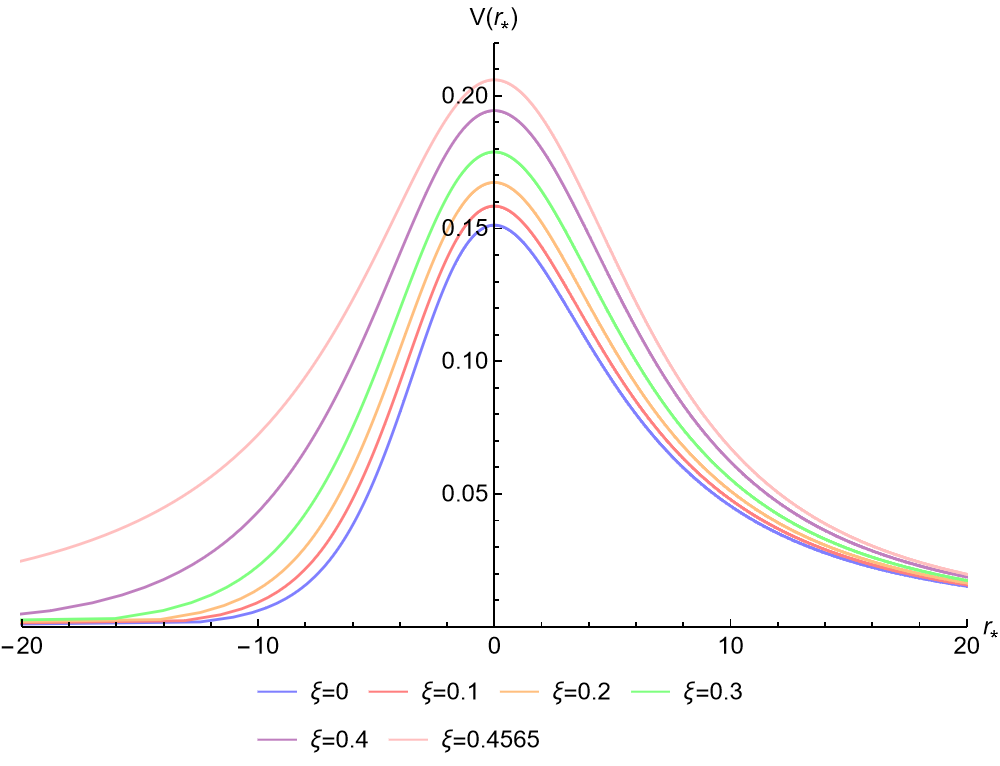}
        \label{fig:l=2vrstar}
    \end{minipage}
    \caption{The effective potential $V (r_{*})$ of the axial gravitational perturbations as a function of the tortoise coordinate for different $l$ with $\xi$ = 0.2 (left panel) and for different $\xi$ with $l = 2$ (right panel).}  
    \label{fig:vrstar}
\end{figure*}


The canonical form for the axial gravitational perturbations in the Regge-Wheeler gauge is given as \cite{Regge:1957td}
{\small
\begin{equation}
    h_{\mu \nu}^{\rm axial}=\left[\begin{array}{cccc}
        0 & 0 & -\csc\theta  \partial_\phi h_0 & \sin\theta  \partial_\theta h_0 \\
        0 & 0 & -\csc\theta  \partial_\phi h_1 & \sin\theta  \partial_\theta h_1 \\
        -\csc\theta  \partial_\phi h_0 & -\csc\theta  \partial_\phi h_1 & 0 & 0 \\
        \sin\theta  \partial_\theta h_0 & \sin\theta  \partial_\theta h_1 & 0 & 0 \\
        \end{array}\right],
\end{equation}
}where $h_{0}(t,r,\theta,\phi)$ and $h_{1}(t,r,\theta,\phi)$ are two unknown functions.
Further, we impose the separable solutions

\begin{equation}
    \begin{split}
        & h_{0}(t,r,\theta ,\phi)=e^{-i  \omega t} \sum_{\ell = 2}^{\infty} \sum_{m=-\ell}^{\ell} h_{0}(r)Y_{\ell m}(\theta ,\phi),\\
        &h_{1}(t,r,\theta ,\phi)=e^{-i  \omega t} \sum_{\ell = 2}^{\infty} \sum_{m=-\ell}^{\ell} h_{1}(r)Y_{\ell m}(\theta ,\phi),
    \end{split}
\end{equation}
where $Y_{\ell m}(\theta ,\phi)$ denotes the spherical harmonic function.




The effective stress-energy tensor corresponding to the BH solution \eqref{bh solution} can be described by that of an anisotropic fluid:
\begin{equation}
    \label{deltaT}
    T_{\nu}^{ \mu}=\rho u^{\mu}u_{\nu}+p_{r} k^{\mu} k_{\nu} +p_{t}\Pi _{\nu}^{ \mu}.
\end{equation}
Here $\rho$, $p_{r}$, $p_{t}$ are the fluid energy density, radial, and tangential pressure, respectively. $k^{\mu}$ is a unit spacelike vector orthogonal to $u^{\mu}$, such that $u^{\mu} u_{\mu}=-1$, $k^{\mu} k_{\mu}=1$, $u^{\mu} k_{\mu}=0$ and $\Pi ^{\mu}_{\nu}=g^{\mu}_{\nu}+u^{\mu} u_{\nu}-k^{\mu} k_{\nu}$. The quantities $\rho$, $p_r$, and $p_t$ are functions of the diagonal components of a rank-2 tensor. Under rotational transformation, their perturbations remain invariant and thus correspond to polar perturbations.


Following \cite{Cardoso:2022whc} the perturbed fluid's 4-velocity components associated with axial perturbations are only given by
\begin{equation}
    \label{deltau}
    \begin{split}
        & \delta u^{\theta} = -\frac{\sqrt{f(r)} \csc \theta }{4\pi \kappa(r) r^{2}} \sum_{\ell = 2}^{\infty} \sum_{m=-\ell}^{\ell} e^{-i  \omega t}U_{\ell m}(r) \partial _{\phi } Y_{\ell m}(\theta,\phi),\\
        &\delta u^{\phi } = \frac{\sqrt{f(r)} \csc^{3} \theta }{4\pi \kappa(r) r^{2}} \sum_{\ell = 2}^{\infty} \sum_{m=-\ell}^{\ell} e^{-i  \omega t}U_{\ell m}(r) \partial _{\theta } Y_{\ell m}(\theta,\phi),
    \end{split}
\end{equation}
where $u^{\mu}=u_{(0)}^{\mu}+\delta u^{\mu}$, $U_{\ell m}(r)$ is an unknown function and by convention $\kappa (r)=\rho(r) + p_{r}(r)+p_{t}(r)$. It can be shown that $\delta k^\mu$ is only related to polar perturbations. Substituting \eqref{deltau} into \eqref{deltaT}, we can obtain the non-zero components of the axial sector of the perturbed stress-energy tensor $\delta T^{\mu}_{\nu}$.  

The perturbed Einstein field equations can be written as
\begin{equation}
    \label{eqE}
     \delta G^{\mu}_{\nu}-8\pi \delta T^{\mu}_{\nu} = \delta \mathcal{E}^{\mu}_{\nu},
\end{equation}
where $\delta G^{\mu}_{\nu}$ is the perturbed Einstein tensor. Substitute the separated variables into \eqref{eqE} and use the combination of $\delta \mathcal{E} ^{\theta}_{\theta }-\delta \mathcal{E} ^{\phi}_{\phi }/\sin^{2}\theta $ we obtain the equation related to $h_{0}(r)$ and $h_{1}(r)$
\begin{equation}
    \label{h0}
    f(r)\left[\frac{h_{1}(r)f'(r)}{f(r)}+h_{1}'(r) \right]+\frac{i\omega h_{0}(r)}{f(r)}=0.
\end{equation}
Moreover, the component $\delta \mathcal{E}^{r} _{\theta }$ gives
\begin{equation}
    \label{h1}
    \begin{split}
        h_{1}(r) (-f(r) (r^2 f''(r)
        & +\ell ^2+\ell -2 f(r))+r^2 \omega ^2) \\
        & -i r^2 \omega  h_{0}'(r)+2 i r \omega  h_{0}(r)=0.
    \end{split}
\end{equation}
It can be observed that $U_{lm}(r)$, $\rho (r)$, $p_{r}(r)$ and $p_{t}(r)$ do not appear in the equations \eqref{h0} and \eqref{h1}, though they persist in other components of the perturbation equations.


After simple algebra manipulation and introducing new variables
\begin{equation}
    h_{1}(r)=\frac{r}{f(r)}\Psi, \quad dr_{*}=\frac{dr}{f(r)} ,
\end{equation}
where $r_{*}$ is the tortoise coordinate, we obtain the master equation in the standard Schr\"odinger-like form 
\begin{equation}
    \label{master equation}
    \frac{d^2\Psi }{d {r_{*}} ^2}+[\omega^{2}-V(r)]\Psi=0,
\end{equation}
where the effective potential is given by
\begin{equation}
    \label{Vrofg}
    V(r)=f(r) \left(\frac{\ell (\ell +1)}{r^2}-\frac{f'(r)}{r}+f''(r)\right)
\end{equation}
where $\ell $ = 2, 3, 4, $\dots$ are the multipole numbers.
It is important to note that the effective potential applies to an anisotropic fluid with an arbitrary non-zero radial pressure.\footnote{In contrast, the effective potential in \cite{Cardoso:2022whc} was obtained specifically for the case of vanishing background radial pressure. In the limit of $p_r^{(0)}=0$, the $rr$-component of the Einstein field equations yields the following relation for the metric gradient: $a'(r)=\frac{a(r)}{r}\frac{1-b(r)}{b(r)}$. Using this relation, Eq. (\ref{Vrofg}) consistently  reproduces the results presented in \cite{Cardoso:2022whc} when the limit $p_r^{(0)} = 0$ is taken.} Note that Ref. \cite{Lutfuoglu:2025ohb} also investigated the axial gravitational perturbations of the BH solution (\ref{bh solution}) and derived a master equation with a different effective potential. This discrepancy arises because our analysis considers general perturbations of the anisotropic fluid, whereas Ref. \cite{Lutfuoglu:2025ohb} restricts the fluid perturbations to the special case $\delta k^{\mu} = \delta u^{\phi} = 0$. Consequently, the resulting effective potentials in the two works differ.

The effective potential $V(r_{*})$ as a function of the tortoise coordinate for different values of the deviation parameter $\xi$ and multipole number $\ell$ is shown in Fig.~\ref{fig:vrstar}. The effective potential is found to be always positive, indicating stability of the BH under axial gravitational perturbations. Moreover, for fixed $\xi$, the height of the potential barrier increases with $\ell$, which is consistent with the behavior observed in both Schwarzschild and Reissner-Nordstr\"om BHs. Similarly, the peak of the effective potential increases with $\xi$ slightly. Importantly, the deviation of the effective potential from that of the Schwarzschild BH becomes particularly significant near the event horizon as $\xi$ increases. This feature may significantly affect the spectrum of the QNMs.

\begin{figure*}  
    \centering
    \begin{minipage}[t]{0.48\textwidth}
        \centering
        \includegraphics[width=\linewidth]{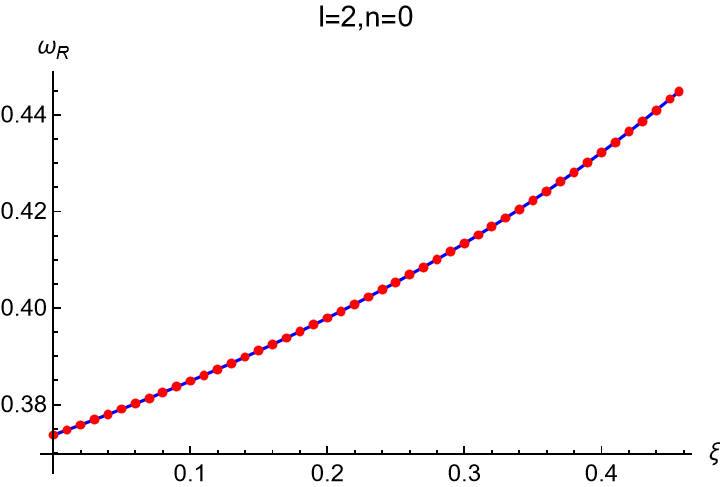}
        \label{fig:Re l=2 n=0}
    \end{minipage}
    \hfill  
    \begin{minipage}[t]{0.48\textwidth}
        \centering
        \includegraphics[width=\linewidth]{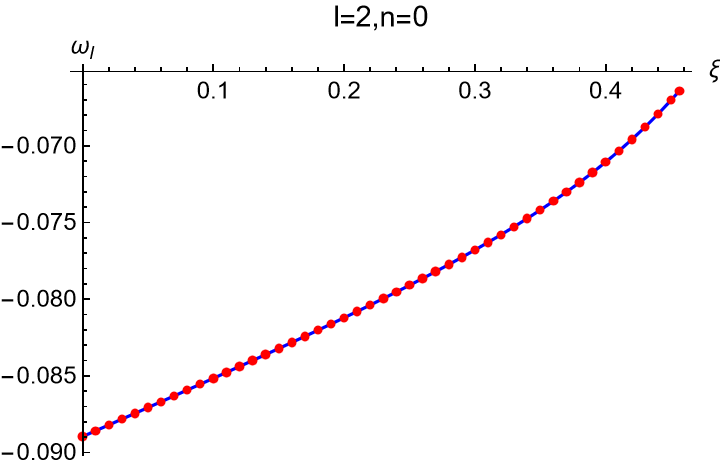}
        \label{fig:Im l=2 n=0}
    \end{minipage}
    \caption{Complex  frequencies of the fundamental $n=0, \ell=2$ mode as a function of the parameter $\xi$. }  
    \label{fig:l=2 n=0}
\end{figure*}

\section{methods}
\label{methods}
The QNM frequency $\omega$ is defined as the eigenvalue of the master equation \eqref{master equation} whose wave solutions must satisfy proper boundary conditions at event horizon and spatial infinity. To be specific, the wave solution of the master equation exhibit pure ingoing behavior at the event horizon and pure outgoing behavior at spatial infinity. That is 
\begin{eqnarray}
&&\Psi\sim e^{-i\omega({r_*}+t)},\quad r\to r_{h},\\
&&\Psi\sim e^{i\omega({r_*}-t)},\quad r\to \infty.
\end{eqnarray}
To obtain the QNMs, with particular attention to higher overtones of axial gravitational perturbations, we employ two commonly used methods to solve the master equation. One is the Bernstein spectral method (BPS) \cite{Fortuna:2020obg}, and the other is the asymptotic iteration method (AIM) \cite{Cho:2009cj,Cho:2011sf}. These two methods can be used to cross-check each other, ensuring the reliability of the results. 
\subsection{Bernstein spectral method}
 In the numerical implementation, it is convenient to work in the ingoing Eddington-Finkelstein (EF) coordinates $(v, r, \theta, \phi)$, where $v=t+r_*$.  Then from the separation of variables 
\begin{equation}
 e^{-i\omega v}\psi(r)=e^{-i\omega t}\Psi(r),
\end{equation}
the master equation \eqref{master equation} in the ingoing EF coordinates becomes
\begin{equation}
    \begin{split}
        \label{master equation2}
        V(r) \psi(r) - f(r) \Bigl[ & \left(-2i\omega + \frac{df(r)}{dr} \right) \frac{d\psi(r)}{dr} \\
        & + f(r) \frac{d^{2}\psi(r)}{dr^{2}} \Bigr]  = 0,
    \end{split}
\end{equation}
with the boundary conditions now being 
\begin{equation}
    \psi\sim {\rm const},\quad  r\to r_{h} \quad {\rm and} \quad \psi\sim e^{2i\omega{r_{*}}}, \quad r\to \infty.
\end{equation}
Furthermore, we perform the following transformation
\begin{equation}
    r=\frac{r_h}{z}.
\end{equation}
Then $z=0$ and $z=1$ correspond to the boundary and the horizon, respectively. Through the following transformation
\begin{equation}
    \label{change}
    \psi(z)=\frac{1}{z} e^{\frac{2i\omega}{z}} z^{-4i\omega}\Phi (z) ,
\end{equation}
the boundary conditions are expressed in the new radial coordinate as 
\begin{equation}
\Phi \sim {\rm const}, \quad z \to 0\quad {\rm and} \quad \Phi \sim {\rm const},\quad z \to 1.
\end{equation}
Substituting \eqref{change} into \eqref{master equation2}, the master equation becomes
\begin{equation}
    \Phi''(z)=\lambda _{0}(z) \Phi'(z)+s _{0}(z)\Phi(z),
\end{equation}
where prime denotes the derivative with respect to $z$, the coefficients $\lambda _{0}(z)$ and $s _{0}(z)$ are given by
{\small
\begin{align}
    \label{lambda0}
    \lambda _0(z) &= \Biggl\{2i\left(3 \xi  z^2 \left(24 \xi  \omega  z^4+z \left(4 \omega +6 \xi  \omega  z^2-3 i z\right)+\omega \right)\right. \nonumber\\[1ex]
    & \left.-\left(6 \xi  z^3+1\right) (2 \omega +z (4 \omega -i)) \log \left(6 \xi  z^3+1\right)\right)\Biggr\} \nonumber\\[1ex]
    & / z^2 \left(6 \xi  z^3+1\right) \left(3 \xi  z^2-\log \left(6 \xi  z^3+1\right)\right)
\end{align}
}

{\small
\begin{align}
    \label{s0}
    s _0(z) & 
        = \Biggl\{\left(3 \xi  z^2-\log \left(6 \xi  z^3+1\right)\right)\left(3z^3\xi \left(\ell  z \left(6 \xi  z^3+1\right)^2\right.\right. \nonumber\\[1ex]
          &+z \left(6 \ell  \xi  z^3+\ell \right)^2+576 \xi ^2 \omega ^2 z^7-2 i \omega \nonumber\\[1ex]
          &+48 \xi  \omega  z^4 \left(4 \omega -3 i \xi  z^3+6 \xi  \omega  z^2-3 i z\right) \nonumber\\[1ex]
          &+4 \left(2 \omega ^2-6 i \xi  \omega  z^3-3 z^2-4 i \omega  z\right) \nonumber\\[1ex]
          &+4 z\left(9 \xi  z^4-30 i \xi  \omega  z^3-6 i \omega  z\right. \nonumber\\[1ex]
          &\left.\left.4 \omega ^2-18 i \xi ^2 \omega  z^5+24 \xi  \omega ^2 z^2\right)\right)\nonumber\\[1ex]
          &-2\left(6 \xi  z^3+1\right)^2\left(2 \omega ^2+4 \omega  z (2 \omega -i)\right. \nonumber\\[1ex]
          &\left.\left.+z^2 \left(8 \omega ^2-6 i \omega -1\right)\right)\log \left(6 \xi  z^3+1\right)\right)\Biggr\} \nonumber\\[1ex]
          &/ z^4 \left(6 \xi  z^3+1\right)^2 \left(\log \left(6 \xi  z^3+1\right)-3 \xi  z^2\right)^2,
\end{align}
}

Due to the presence of the logarithmic term $\log (6 \xi z^3+1)$, which diverges as $z \to 0$, numerical instabilities arise in the Bernstein spectral method, obstructing the computation of QNMs. To address this issue while maintaining computational precision, we expand the logarithm as:
\begin{equation}
    \log(6\xi z^3 + 1) = \sum_{a=1}^{b} (-1)^{a+1} \frac{6^a \xi^a z^{3a}}{a} + \mathcal{O}(\xi^{b+1})
\end{equation}
where the truncation order $b$ is dynamically chosen within the range of 14 to 20. This specific range optimizes the trade-off between accuracy and stability: lower-order expansions ($b<14$) result in insufficient precision for QNMs  calculation, while higher-order expansions ($b>20$) reintroduce divergent behavior.

We represent $\Phi (z)$ as a sum

\begin{equation}
    \label{sum}
    \Phi(z)=\sum_{k=0}^{N} C_{k}B_{k}^{N}(z),
\end{equation}
where

\begin{equation}
    B_{k}^{N}(z)\equiv \frac{N!}{k!(N-k)!} z^{k}(1-z)^{N-k}.
\end{equation}
are the Bernstein polynomials. Using a Chebyshev collocation grid of N + 1 points,

\begin{equation}
z_{p}=\frac{1-\cos \frac{p \pi }{N} }{2} =\sin^{2}\frac{p \pi }{2N}, p \in \{0, 1, 2, \dots, N\}  ,
\end{equation}
we derive a system of linear equations for the coefficients $C_k$. Non-trivial solutions exist if and only if the corresponding coefficient matrix (whose elements are polynomials of $\omega$) becomes singular. This condition reduces the problem to a numerical generalized eigenvalue problem concerning the $\omega$. Solving this eigenvalue problem numerically yields the eigenfrequencies $\omega$ and allows for the subsequent calculation of the coefficients $C_k$. These coefficients explicitly define the polynomial \eqref{sum}, which approximates the solution to the wave equation.

To eliminate spurious eigenvalues arising from the finite polynomial basis in \eqref{sum}, we implemented comparative analyses with varying grid resolutions N. As detailed in \cite{Fortuna:2020obg}, their computational package provides systematic procedures for such multi-resolution comparisons. By progressively increasing N, we simultaneously enhanced numerical precision, thereby improving confidence in eigenvalue convergence. It is worth noting that in certain scenarios, the Bernstein spectral method demonstrates superiority over other basis functions (including Chebyshev and Fourier) in terms of numerical cost or solution accuracy \cite{Fortuna:2020obg}. We performed explicit comparisons with Chebyshev-basis spectral methods and confirmed this conclusion. Specifically, for $\ell =2$ configurations approaching $\xi \to \xi _{cr}$, the Chebyshev-based spectral method could only resolve modes with $n=0$ and $n=1$, whereas the Bernstein-based approach successfully captured the first six overtones.

\begin{figure*}  
    \centering
    \begin{minipage}[t]{0.48\textwidth}
        \centering
        \includegraphics[width=0.9\textwidth]{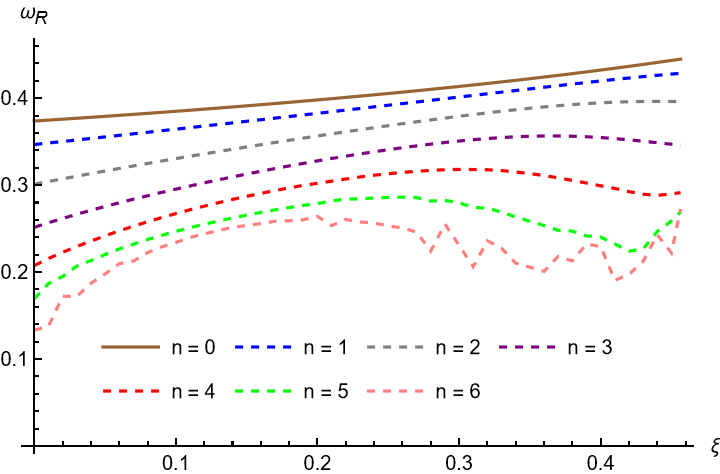}
        \label{Re l=2}
    \end{minipage}
    \hfill  
    \begin{minipage}[t]{0.48\textwidth}
        \centering
        \includegraphics[width=0.9\textwidth]{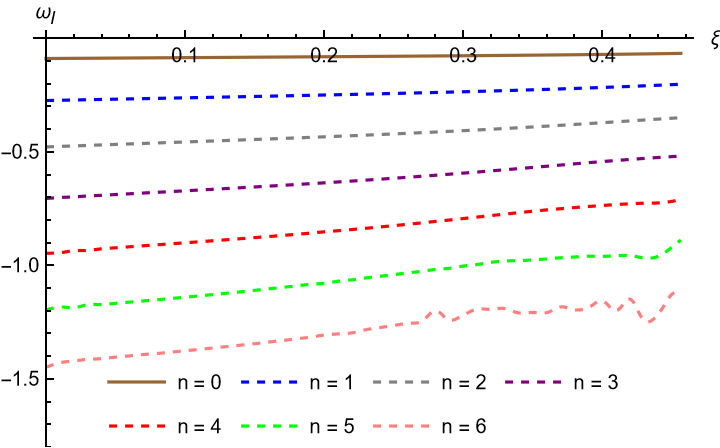} 
        \label{Im l=2}
    \end{minipage}
    \caption{$\text{Re}(\omega)$ (left panel) and $\text{Im}(\omega)$ (right panel) as  functions of $\xi$ for the fundamental mode and first six overtones (top to bottom) with $\ell=2$.}  
    \label{fig:l=2AIM}
\end{figure*}

\begin{figure*}  
    \centering
    \begin{minipage}[t]{0.48\textwidth}
        \centering
        \includegraphics[width=0.9\textwidth]{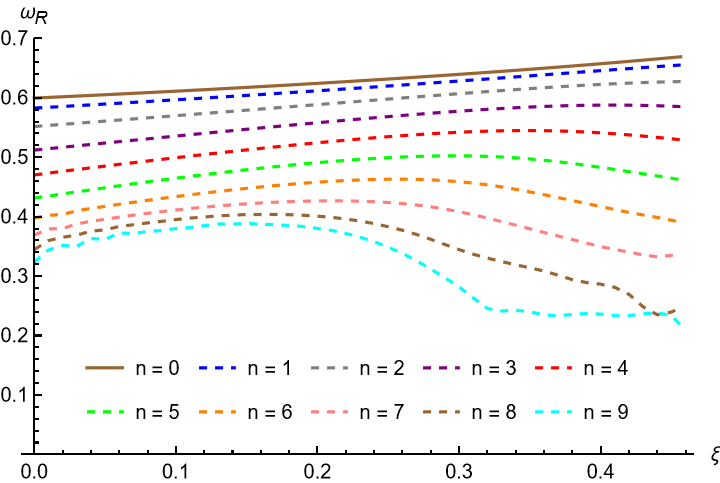}
        \label{Re l=3}
    \end{minipage}
    \hfill  
    \begin{minipage}[t]{0.48\textwidth}
        \centering
        \includegraphics[width=0.9\textwidth]{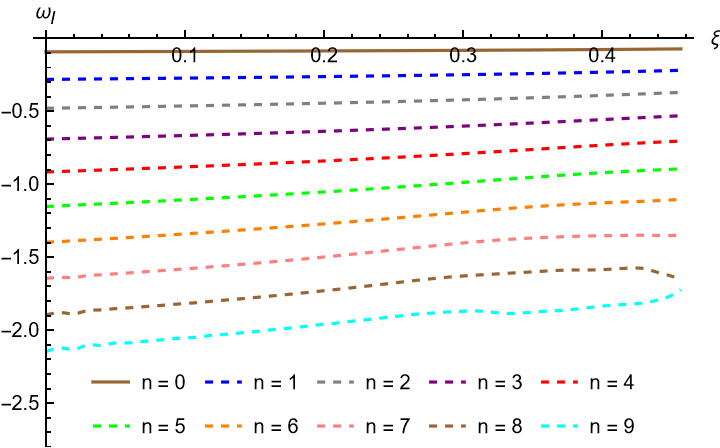} 
        \label{Im l=3}
    \end{minipage}
    \caption{$\text{Re}(\omega)$ (left panel) and $\text{Im}(\omega)$ (right panel) as  functions of $\xi$ for the fundamental mode and first nine overtones (top to bottom) with $\ell=3$.}  
    \label{fig:l=3AIM}
\end{figure*}

\begin{table*}[t]
    \centering
    \captionsetup{justification=raggedright, singlelinecheck=false} 
    \caption{
Frequencies of QNMs with $\ell = 2$, $n = 0$ calculated using different methods for various values of $\xi$. Differences are shown relative to the Schwarzschild BH, along with the discrepancy between the methods. An error of $\SI{0.00}{\percent}$ indicates that the difference is smaller than $10^{-7}$.
}
    \begin{tabular}{lccccc}
         \hline
        $\xi$ & $\omega (\text{BPS})$ & $\omega (\text{AIM})$ & $\Delta \omega_{R}$ & $\Delta \omega_{I}$ & $\Delta \omega$ \\
        \hline
        0.00  & 0.373672 - 0.0889623i & 0.373672 - 0.0889623i & $\SI{0.00}{\percent}$ & $\SI{0.00}{\percent}$ & $\SI{0.00}{\percent}$ \\
        \addlinespace
        0.05  & 0.379058 - 0.0870715i & 0.379058 - 0.0870715i & $\SI{1.4414}{\percent}$ & $\SI{2.1251}{\percent}$ & $\SI{0.00}{\percent}$\\
        \addlinespace
        0.10  & 0.384864 - 0.0851638i & 0.384864 - 0.0851638i & $\SI{2.9951}{\percent}$ & $\SI{4.2695}{\percent}$ & $\SI{0.00}{\percent}$\\
        \addlinespace
        0.15  & 0.391135 - 0.0832211i & 0.391135 - 0.0832211i & $\SI{4.6734}{\percent}$ & $\SI{6.4532}{\percent}$ & $\SI{0.00}{\percent}$\\
        \addlinespace
        0.20  & 0.397928 - 0.0812138i & 0.397928 - 0.0812139i & $\SI{6.4913}{\percent}$ & $\SI{8.7095}{\percent}$ & $\SI{0.00001}{\percent}$\\
        \addlinespace
        0.25  & 0.405317 - 0.0790961i & 0.405317 - 0.0790962i & $\SI{8.4687}{\percent}$ & $\SI{11.0900}{\percent}$ & $\SI{0.00001}{\percent}$\\
        \addlinespace
        0.30  & 0.413402 - 0.0767946i & 0.413402 - 0.0767946i & $\SI{10.6323}{\percent}$ & $\SI{13.6771}{\percent}$ & $\SI{0.00}{\percent}$\\
        \addlinespace
        0.35  & 0.422314 - 0.0741881i & 0.422314 - 0.074188i & $\SI{13.0173}{\percent}$ & $\SI{16.6070}{\percent}$ & $\SI{0.00001}{\percent}$\\
        \addlinespace
        0.40  & 0.432224 - 0.0710653i & 0.432225 - 0.0710639i & $\SI{15.6695}{\percent}$ & $\SI{20.1180}{\percent}$ & $\SI{0.00019}{\percent}$\\
        \addlinespace
        0.45  & 0.443306 - 0.0668701i & 0.443316 - 0.0670279i & $\SI{18.6364}{\percent}$ & $\SI{24.7443}{\percent}$ & $\SI{0.01763}{\percent}$\\
        \addlinespace
        0.4565 & 0.444842 - 0.0664069i & 0.444905 - 0.067279i & $\SI{19.0545}{\percent}$ & $\SI{24.8635}{\percent}$ & $\SI{0.09718}{\percent}$\\
        \hline
    \end{tabular}
    \label{tab:l2_n0}
\end{table*}

\subsection{Asymptotic iteration method}

The AIM is a semi-analytical method for solving second-order ordinary differential equations. It is particularly well-suited for dealing with eigenvalue problems, such as the calculation of QNMs in quantum mechanics and gravitational perturbations \cite{Cho:2009cj,Cho:2011sf}.

Let us consider a second-order differential equation of the form: 

\begin{equation}
    \label{AIM01}
\chi''(x) = \lambda_0(x) \chi'(x) + s_0(x) \chi(x),
\end{equation}
where prime denotes the derivative with respect to $x$, $\lambda_0(x)$ and $s_0(x)$ are well-defined functions and sufficiently smooth. To find a general solution to this equation, we differentiate \eqref{AIM01} with respect to $\chi$ and find that

\begin{equation}
\chi''' = \lambda_1(x) \chi' + s_1(x) \chi,
\end{equation}
where the new two coefficients are given by:
\begin{equation}
\lambda_1(x) = \lambda_0' + s_0 + \lambda_0^2, \quad s_1(x) = s_0' + s_0 \lambda_0.
\end{equation}
Using this process iteratively, we obtain the $(n + 2)$-th derivative of $\chi(x)$:

\begin{equation}
\chi^{(n+2)} = \lambda_n(x) \chi' + s_n(x) \chi,
\end{equation}
where the new coefficients $\lambda_n(x)$ and $s_n(x)$ are related to the lower-order's ones as follows:

\begin{equation}
    \label{iteration}
\lambda_n(x) = \lambda_{n-1}' + s_{n-1} + \lambda_0 \lambda_{n-1}, \quad 
s_n(x) = s_{n-1}' + s_0 \lambda_{n-1}.
\end{equation}
For sufficiently large values of $n$, the method demonstrates asymptotic characteristics - through iterative refinement, the system ultimately converges to a stable oscillation mode (corresponding to the BH QNMs), as specified below.

\begin{equation}
\frac{s_n(x)}{\lambda_n(x)} = \frac{s_{n-1}(x)}{\lambda_{n-1}(x)} = {\rm const}.
\end{equation}
The perturbation frequencies can be obtained from the above ``quantization condition''. However, this procedure has a difficulty: the process of taking the derivative of $\lambda_n(x)$ and $s_n(x)$ at each iteration step consumes much time and affects numerical precision.  To circumvent these issues, $\lambda_n(x)$ and $s_n(x)$ are expanded in Taylor series around the point $x$ at which the AIM method is performed:

\begin{equation}
\lambda_n(x) = \sum_{i=0}^{\infty} c_i^n (x - x_0)^i, \quad 
s_n(x) = \sum_{i=0}^{\infty} d_i^n (x - x_0)^i,
\end{equation}
where $c_i^n$ and $d_i^n$ are the $i$-th Taylor coefficients of $\lambda_n(x)$ and $s_n(x)$, respectively. Substituting the above equations into \eqref{iteration} leads to a set of recursion relations for the Taylor coefficients:

\begin{equation}
c_i^n = \sum_{k=0}^{i} c_k^0 c_{i-k}^{n-1} + (i+1) c_{i+1}^{n-1} + d_i^{n-1},
\end{equation}

\begin{equation}
d_i^n = \sum_{k=0}^{i} d_k^0 c_{i-k}^{n-1} + (i+1) d_{i+1}^{n-1}.
\end{equation}

In terms of these coefficients, the “quantization condition” equation can be reexpressed as:

\begin{equation}
d_0^n c_0^{n-1} - c_0^n d_0^{n-1} = 0,
\end{equation}

Now we have simplified the AIM into a set of recursive relations that do not require derivative operators. For our work, we write the obtained perturbation equation in the form of equation \eqref{AIM01}, where the coefficients $\lambda_0(z)$ and $s_0(z)$ are given by \eqref{lambda0} and \eqref{s0}. For the AIM, we choose the same coordinate transformation \( z = \frac{r_h}{r} \) as that when using the Bernstein spectral method.

\section{QUASINORMAL MODES}
\label{QUASINORMAL MODES}
\subsection{Fundamental mode}
 The QNMs of various test fields in the background of the regular BH (\ref{bh solution}) was studied in \cite{Stashko:2024wuq}, where the author found that the fundamental modes are only weakly sensitive to the scale parameter  $\xi$. In this subsection, we will further investigate the general properties of the fundamental modes arising from the axial gravitational perturbations in the same spacetime.

Fig.\ref{fig:l=2 n=0} illustrates the real and imaginary part of the $l = 2$ and $n = 0$ fundamental mode  as functions of the parameter $ \xi $. Here $n$ denotes the overtone number and is defined as an integer labeling the QNMs by increasing the decay rate $|\text{Im}(\omega)|$. We observe that both the real part $\text{Re}(\omega)$ and the imaginary part $\text{Im}(\omega)$  increase monotonically  with $\xi$, consistent with the behavior observed for massless test fields in \cite{Stashko:2024wuq}. This indicates that the oscillation period and the damping rate decreases as $\xi$ increases. Notably, the deviation between the regular BH and the Schwarzschild solution becomes more pronounced with increasing  $\xi$. As the extremal limit is approached  ($ \xi \to \xi_{\text{cr}} $), the maximum relative deviations in $\text{Re}(\omega)$ and $\text{Im}(\omega)$ compared to the Schwarzschild case reach approximately  $\SI{19}{\percent}$ and $\SI{25}{\percent}$, respectively.

Comparing our results with those in \cite{Lutfuoglu:2025ohb} under the condition $l = 2$ and $n = 0$, it can be observed that the QNMs obtained using the effective potential (\ref{Vrofg}) have a larger value of real part and a smaller absolute value of imaginary part. Furthermore, as $\xi$ increases, the variation of QNMs becomes more pronounced. The reason for this discrepancy may lie in the different perturbation ansatz and treatment of fluid degrees of freedom.

Subsequently, we compute the frequency of the QNMs for $\ell \geq 3$ with $n = 0$, and find that both  $\text{Re}(\omega)$ and $\text{Im}(\omega)$ exhibit the same pronounced monotonic dependence on $\xi$ as observed in the $\ell = 2$ case.
The fundamental modes for $\ell = 2$, calculated using both methods, are explicitly listed in Tab.~\ref{tab:l2_n0}. We also evaluate the relative deviations of the real and imaginary parts of the frequencies from those of the Schwarzschild BH, defined as
\begin{equation}
\Delta \omega_R = \frac{\omega_R^{\text{(AIM)}} - \omega_R^{\text{(SBH)}}}{\omega_R^{\text{(SBH)}}}, \quad
\Delta \omega_I = \frac{\omega_I^{\text{(SBH)}} - \omega_I^{\text{(AIM)}}}{\omega_I^{\text{(SBH)}}}.
\end{equation}
Furthermore, a quantitative comparison between the two numerical methods (AIM and BPS) is provided using the relative difference,
\begin{equation}
\Delta \omega = \frac{\left| \omega^{\text{(AIM)}} - \omega^{\text{(BPS)}} \right|}{\left| \omega^{\text{(AIM)}} + \omega^{\text{(BPS)}} \right|}.
\end{equation}
From Tab.~\ref{tab:l2_n0} we know that except for the case approaching the extreme BH, the results obtained from the two different methods are in excellent agreement.
\begin{table*}[t]
    \captionsetup{justification=raggedright, singlelinecheck=false} 
    \caption{The list of $\omega$, $\Delta \omega_{R}$ and $\Delta \omega_{I}$ of the QNMs for different overtones $n$ and $\xi$ with $\ell = 2$.}
\begin{tabular}{lcccc}
    \hline
    $n$ & $\omega (\xi=0)$ & $\omega (\xi=0.2)$ & $\Delta \omega_{R}$ & $\Delta \omega_{I}$\\
    \hline
    0 & 0.373672 - 0.088962i & 0.397928 - 0.081214i & $\SI{6.4913}{\percent}$ & $\SI{8.7095}{\percent}$ \\
    \addlinespace
    1 & 0.346711 - 0.273915i & 0.382549 - 0.249611i & $\SI{10.3366}{\percent}$ & $\SI{8.8728}{\percent}$ \\
    \addlinespace
    2 & 0.301053 - 0.478277i & 0.356523 - 0.433733i & $\SI{18.4253}{\percent}$ & $\SI{9.3134}{\percent}$ \\
    \addlinespace
    3 & 0.251505 - 0.705148i & 0.327926 - 0.636398i & $\SI{30.3855}{\percent}$ & $\SI{9.7497}{\percent}$ \\
    \addlinespace
    4 & 0.207515 - 0.946844i & 0.302115 - 0.852079i & $\SI{45.5871}{\percent}$ & $\SI{10.0085}{\percent}$ \\
    \addlinespace
    5 & 0.169315 - 1.195600i & 0.278746 - 1.078901i & $\SI{64.6316}{\percent}$ & $\SI{9.7607}{\percent}$ \\
    \addlinespace
    6 & 0.133287 - 1.448116i & 0.264062 - 1.306830i & $\SI{98.1153}{\percent}$ & $\SI{9.7565}{\percent}$ \\
    \hline
\end{tabular}
    \label{tab:l2}
\end{table*}

\begin{table*}[t]
    \captionsetup{justification=raggedright, singlelinecheck=false} 
    \caption{The list of $\omega$, $\Delta \omega_{R}$ and $\Delta \omega_{I}$ of the QNMs for different overtones $n$ and $\xi$ with $\ell = 3$. }
\begin{tabular}{lcccc}
    \hline
    $n$ & $\omega (\xi=0)$ & $\omega (\xi=0.2)$ & $\Delta \omega_{R}$ & $\Delta \omega_{I}$ \\
    \hline
    0 & 0.599443 - 0.092703i & 0.624219 - 0.086564i & $\SI{4.1332}{\percent}$ & $\SI{6.6222}{\percent}$ \\
    \addlinespace
    1 & 0.582644 - 0.281298i & 0.611731 - 0.262100i & $\SI{4.9922}{\percent}$ & $\SI{6.8248}{\percent}$ \\
    \addlinespace
    2 & 0.551685 - 0.479093i & 0.588526 - 0.444485i & $\SI{6.6779}{\percent}$ & $\SI{7.2237}{\percent}$ \\
    \addlinespace
    3 & 0.511962 - 0.690337i & 0.558006 - 0.636833i & $\SI{8.9936}{\percent}$ & $\SI{7.7504}{\percent}$ \\
    \addlinespace
    4 & 0.470174 - 0.915649i & 0.524320 - 0.839980i & $\SI{11.5162}{\percent}$ & $\SI{8.2640}{\percent}$ \\
    \addlinespace
    5 & 0.431386 - 1.152151i & 0.490824 - 1.052265i & $\SI{13.7784}{\percent}$ & $\SI{8.6695}{\percent}$ \\
    \addlinespace
    6 & 0.397660 - 1.395912i & 0.458002 - 1.271310i & $\SI{15.1743}{\percent}$ & $\SI{8.9262}{\percent}$ \\
    \addlinespace
    7 & 0.368992 - 1.643845i & 0.426445 - 1.497981i & $\SI{15.5703}{\percent}$ & $\SI{8.8733}{\percent}$ \\
    \addlinespace
    8 & 0.344619 - 1.894032i & 0.400812 - 1.729105i & $\SI{16.3058}{\percent}$ & $\SI{8.7077}{\percent}$ \\
    \addlinespace
    9 & 0.323682 - 2.145396i & 0.380094 - 1.958998i & $\SI{17.4282}{\percent}$ & $\SI{8.6883}{\percent}$ \\
    \hline
\end{tabular}
\label{tab:l3}
\end{table*}

\subsection{Higher overtones}

The AIM and Bernstein spectral methods yield consistent results for low overtone quasinormal modes. However, at higher overtones, the Bernstein spectral method becomes highly sensitive to the number of grid points: varying the grid-point density leads to significant fluctuations in the computed frequencies. In contrast, the AIM demonstrates stable performance across different iteration counts. Therefore, in all subsequent figures and tables, we present results exclusively obtained using the AIM.

As previously observed, the actual oscillation frequency of fundamental modes monotonically increases with the parameter $\xi$. However, the behavior changes for higher overtones. Fig.\ref{fig:l=2AIM} shows $\text{Re}(\omega)$ and $\text{Im}(\omega)$ as  functions of $\xi$ for the fundamental mode and first six overtones with $\ell=2$. From this figure, we can observe that starting from the second overtone, $\text{Re}(\omega)$ no longer exhibits monotonic growth. Specifically, 
$\text{Re}(\omega)$ initially increases with $\xi$ up to a relatively large critical value, beyond which it begins to decrease. Similarly, the damping rate, reflected in the magnitude of $\text{Im}(\omega)$ no longer exhibits monotonic decrease as $\xi$ increases starting the fifth overtone.  A pronounced ``outburst'' phenomenon in $\text{Re}(\omega)$ appears in the fifth and sixth overtones, while a similar anomaly in $\text{Im}(\omega)$ becomes evident in the sixth overtone. This behavior may originate from structural differences near the event horizons between the Schwarzschild BH and the regular BH, as discussed in~\cite{Konoplya:2022hll, Konoplya:2022pbc, Konoplya:2022iyn, Gong:2023ghh}. From Fig.\ref{fig:vrstar}, it is clear that the effective potentials of the regular and Schwarzschild BH differ significantly as functions of the tortoise coordinate for $\ell =2$ and $\ell =3$. These changes in the effective potential with increasing $\xi$ likely contribute to the emergence of overtone outbursts.

We also calculate the first nine overtones for $\ell =3$. As shown in Fig.\ref{fig:l=3AIM}, a clear outburst of $\text{Re}(\omega)$ and a mild outburst of $\text{Im}(\omega)$ are observed  in the ninth overtone, reinforcing the universality of this non-monotonic behavior across different multipole numbers. Tabs.\ref{tab:l2} and Tabs.\ref{tab:l3} list the exact values of frequencies of the fundamental mode and first few overtones for $\ell  = 2$ and $\ell  = 3$, respectively. As the overtone increases, the deviations in both $\text{Re}(\omega)$ and $\text{Im}(\omega)$ from the Schwarzschild case become more pronounced, further indicating that higher overtone modes are more sensitive to variations in $\xi$ than the fundamental ones.

By comparing the overtone behavior of axial gravitational perturbations with that of test field perturbations~\cite{Stashko:2024wuq}, we find that the outburst phenomenon becomes increasingly prominent at relatively low values of the angular momentum number $l$. For test fields, the first overtone ($n=1$) already exhibits a clear outburst behavior in the quasinormal frequency spectrum for $l=0$, whereas for $l=1$ this feature appears only for higher overtones, within the range considered up to $n=4$. In contrast, for axial gravitational perturbations the outburst is significantly suppressed and shifted to higher overtone orders. For instance, in the case $l=2$, indications of outburst behavior become visible only when overtones up to at least $n=5$ are taken into account.

\section{GREY-BODY FACTORS and THE CORRESPONDENCE WITH QNMs}
\label{GREY-BODY FACTORS OBTAINED VIA THE CORRESPONDENCE WITH QUASINORMAL MODES}

\begin{figure*}  
    \centering
    \begin{minipage}[t]{0.48\textwidth}
        \centering
        \includegraphics[width=\linewidth]{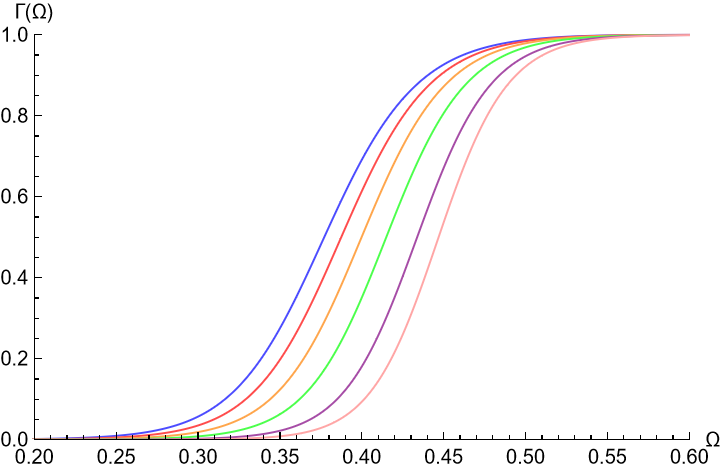}
        \label{fig:l=2gama}
    \end{minipage}
    \hfill  
    \begin{minipage}[t]{0.48\textwidth}
        \centering
        \includegraphics[width=\linewidth]{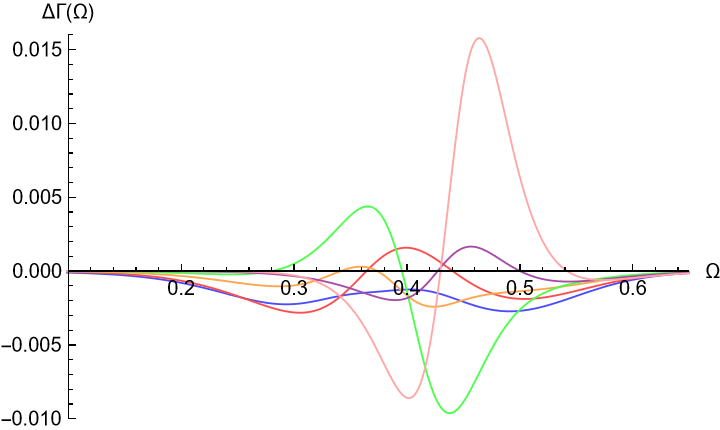}
        \label{fig:l=2deltagama}
    \end{minipage}
    \caption{Left panel: Grey-body factors of the $l=2$ axial gravitational perturbations of the regular BH for $\xi$ = 0 (blue), $\xi$ = 0.1 (red), $\xi$ = 0.2 (orange), $\xi$ = 0.3 (green), $\xi$ = 0.4 (purple), $\xi$ = 0.4565 (pink), $M$ = 1. Right panel: difference between grey-body factors obtained by the 6th order WKB and the correspondence. }  
    \label{fig:graybodyl2}
\end{figure*}

\begin{figure*}  
    \centering
    \begin{minipage}[t]{0.48\textwidth}
        \centering
        \includegraphics[width=\linewidth]{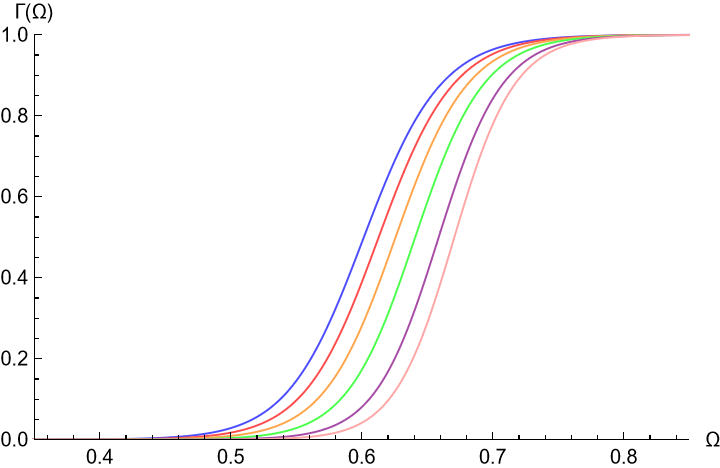}
        \label{fig:l=3gama}
    \end{minipage}
    \hfill  
    \begin{minipage}[t]{0.48\textwidth}
        \centering
        \includegraphics[width=\linewidth]{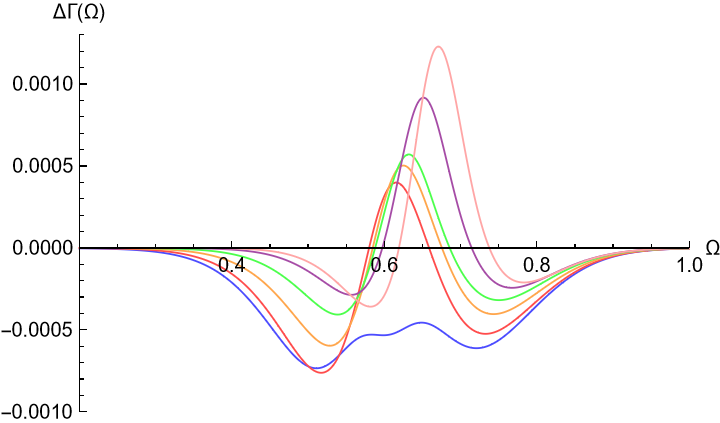}
        \label{fig:l=3deltagama}
    \end{minipage}
    \caption{Left panel: Grey-body factors of the $\ell =3$ axial gravitational perturbations of the regular BH for $\xi$ = 0 (blue), $\xi$ = 0.1 (red), $\xi$ = 0.2 (orange), $\xi$ = 0.3 (green), $\xi$ = 0.4 (purple), $\xi$ = 0.4565 (pink), $M$ = 1. Right panel: difference between grey-body factors obtained by the 6th order WKB and the correspondence. }  
    \label{fig:graybodyl3}
\end{figure*}

 In this section, we investigate the impact of quantum corrections on the grey-body factor and examine the validity of the correspondence between QNMs and grey-body factors in the high-frequency limit, as established for spherically symmetric BHs in \cite{Konoplya:2024lir}. Our analysis focuses on axial gravitational perturbations of the regular BH within the framework of asymptotically safe gravity.
 
 In the scattering process around a BH, the grey-body factors can be used to observe the reflection of waves by potential barriers. When grey-body factors approach 1, they indicate minimal reflection, implying complete transmission of waves through the potential barrier. For a wave scattered by a BH, the boundary conditions have the following form
\begin{equation}
    \Psi = 
    \begin{cases}
        e^{-i \Omega r_{*}} + R e^{i \Omega r_{*}}, &  r_{*} \to +\infty, \\
        T e^{-i \Omega r_{*}},                       &  r_{*} \to -\infty,
    \end{cases}
\end{equation}
where $\Omega$ is the real frequency of the wave, $R$ and $T$ represent the reflection coefficient and transmission coefficient, respectively. The conservation of energy of the wave requires that $|T|^2+|R|^2=1$. In the context of BH radiation, the transmission coefficient $T$ is related to the grey-body factor by
\begin{equation}
    \Gamma _{\ell }(\Omega )= \left | T \right |^{2} =1-\left | R\right |^{2}.
\end{equation}
When the BH spacetime is spherically symmetric, the WKB method allows us to express the  the grey-body factor as follows \cite{Iyer:1986np} 
\begin{equation}
	\Gamma _{\ell }(\Omega )=\frac{1}{1+e^{2 \pi i \mathcal{K} }} ,
\end{equation}
where $\mathcal{K}$ is a function of the frequency $\Omega$, defined through the equation derived in the framework of the WKB   \cite{Konoplya:2019hlu}:
{\small
\begin{equation}\label{WKBeq}
    \begin{aligned}
        \Omega^{2} & = V_{0} + A_{2}\left(\mathcal{K}^{2}\right) + A_{4}\left(\mathcal{K}^{2}\right) + A_{6}\left(\mathcal{K}^{2}\right) + \ldots \\
        & \quad -i \mathcal{K} \sqrt{-2 V_{2}} \left(1 + A_{3}\left(\mathcal{K}^{2}\right) + A_{5}\left(\mathcal{K}^{2}\right) + A_{7}\left(\mathcal{K}^{2}\right) + \ldots \right).
    \end{aligned}
\end{equation}
}
For a given frequency $\Omega$, the solution to can be formally written as
\begin{equation}
    \mathcal{K}=\frac{i\Omega^{2}-V_{0}}{\sqrt{-2 V_{2}}}+\sum_{i=2}^{6} \Lambda_{i}(\mathcal{K}).
\end{equation}
where $V_{0}$ is the peak value of the effective potential, $V_{2}$ denotes the second derivative of the effective potential at the same point, $A_{i}(\mathcal{K}^{2})$ and $\Lambda_{i}(\mathcal{K})$ represents higher-order WKB corrections. For more details about above formulas, one can refer to \cite{Iyer:1986np,Konoplya:2003ii,Matyjasek:2017psv}.

Note that (\ref{WKBeq}) can be used to calculate the complex frequencies of QNMs by replacing $\Omega\to \omega$ and let $\mathcal{K}=n+\frac{1}{2}$, where $n$ is the overtone number. Since both the grey-body factor and QNMs are closely related to the effective potential and are governed by equations of the same form within the WKB approximation, it is natural to expect a connection between these two quantities. Indeed, Konoplya and Zhidenko \cite{Konoplya:2024lir} established a correspondence between grey-body factors and QNMs for spherically symmetric black holes using the WKB method. Further, they studied the correspondence in the rotating BH case \cite{Konoplya:2024vuj}. The correspondence provide an approximate method to calculate the grey-body factor from the QNMs. To be specific, it was found in that $\mathcal{K}$ appearing in the expression of the gray-body factor can be determined by
\begin{equation}\label{correspondenceformula}
    -i \mathcal{K}=-\frac{\Omega^{2}-\operatorname{Re}\left(\omega_{0}\right)^{2}}{4 \operatorname{Re}\left(\omega_{0}\right) \operatorname{Im}\left(\omega_{0}\right)}+\Delta_{1}+\Delta_{2}+\Delta_{f}+\mathcal{O}\left(\ell^{-3}\right),
\end{equation}
where
\begin{equation}
    \Delta _{1}=\frac{Re(\omega _{0})-Re(\omega _{1})}{16Im(\omega _{0})} +\mathcal{O}(\ell^{-2}),
\end{equation}
\begin{equation}
    \begin{aligned}
        \Delta_{2} &= -\frac{\Omega^{2} - \operatorname{Re}(\omega_{0})^{2}}{32 \operatorname{Re}(\omega_{0}) \operatorname{Im}(\omega_{0})}
            \left( \frac{ \left( \operatorname{Re}(\omega_{0}) - \operatorname{Re}(\omega_{1}) \right)^{2} }{4 \operatorname{Im}(\omega_{0})^{2}} \right. \\
        &\quad \left. - \frac{3 \operatorname{Im}(\omega_{0}) - \operatorname{Im}(\omega_{1})}{3 \operatorname{Im}(\omega_{0})} \right)
            + \frac{ \left( \Omega^{2} - \operatorname{Re}(\omega_{0})^{2} \right)^{2} }{16 \operatorname{Re}(\omega_{0})^{3} \operatorname{Im}(\omega_{0})} \\
        &\quad \times \left( 1 + \frac{ \operatorname{Re}(\omega_{0}) \left( \operatorname{Re}(\omega_{0}) - \operatorname{Re}(\omega_{1}) \right) }{4 \operatorname{Im}(\omega_{0})^{2}} \right)
            + \mathcal{O}\left( \ell^{-3} \right),
    \end{aligned}
\end{equation}
and
{\small
\begin{equation}
    \begin{aligned}
        &\Delta_{f} = \frac{\left(-\Omega^{2}+\operatorname{Re}\left(\omega_{0}\right)^{2}\right)^{3}}{32 \operatorname{Re}\left(\omega_{0}\right)^{5} \operatorname{Im}\left(\omega_{0}\right)}
        \left( 1 + \frac{\operatorname{Re}\left(\omega_{0}\right)\left(\operatorname{Re}\left(\omega_{0}\right)-\operatorname{Re}\left(\omega_{1}\right)\right)}{4 \operatorname{Im}\left(\omega_{0}\right)^{2}} \right. \\
        & \left. + \operatorname{Re}\left(\omega_{0}\right)^{2} \left( 
        \frac{\left(\operatorname{Re}\left(\omega_{0}\right)-\operatorname{Re}\left(\omega_{1}\right)\right)^{2}}{16 \operatorname{Im}\left(\omega_{0}\right)^{4}} 
        - \frac{3 \operatorname{Im}\left(\omega_{0}\right)-\operatorname{Im}\left(\omega_{1}\right)}{12 \operatorname{Im}\left(\omega_{0}\right)} \right) \right) \\
        & + \mathcal{O}\left(\ell^{-3}\right).
    \end{aligned}
\end{equation}
}
Here, $\omega _{0} $ and $\omega _{1} $ are the frequencies of the fundamental and first overtone QNMs, respectively.

To verify the QNM/grey-body factor correspondence for axial gravitational perturbations of the regular BH \cite{Bonanno:2023rzk}, we apply the 6th-order WKB approximation \cite{Konoplya:2003ii} to calculate the grey-body factors. We then compare these results with those obtained via the  correspondence by substituting the accurate QNM frequencies obtained in previous section into the formula \ref{correspondenceformula}.
In Figs. \ref{fig:graybodyl2} and \ref{fig:graybodyl3}, we present the grey-body factors computed using the sixth-order WKB approximation as functions of the wave frequency $\Omega$ for $\ell = 2, 3$ and various values of the coupling parameter $\xi$. It is observed that the grey-body factors gradually decrease with increasing $\xi$. This behavior can be attributed to the influence of quantum effects on the effective potential, as illustrated in the right panel of Fig. \ref{fig:vrstar}. As $\xi$ increases, the effective potential becomes higher, leading to a more pronounced potential barrier. This, in turn, reduces the transmission coefficient and consequently lowers the grey-body factors. However, the dependence of the grey-body factors on $\xi$ is relatively mild compared to the overtone variations shown in Fig. \ref{fig:l=2AIM}, indicating a lower sensitivity. As suggested by the correspondence, this is because the grey-body factors are mainly governed by the fundamental mode and the first overtone, both of which exhibit only minor deviations from the Schwarzschild case. This supports the conclusion that the grey-body factors represent a more stable and robust feature of BH radiation.

Moreover, in Figs. \ref{fig:graybodyl2} and \ref{fig:graybodyl3} we compare the grey-body factors calculated using the 6th-order WKB method with those obtained via the correspondence from $\omega_0$ and $\omega_1$ for $\ell =2$ and $\ell =3$. The results indicate that the differences remain within two percent for $\ell =2$ and one thousandth for $\ell =3$, and the accuracy becomes even higher for larger $\ell$.

\section{CONCLUSIONS}
\label{CONCLUSIONS}

In this paper, we present a detailed investigation of axial gravitational perturbations of the regular BH model proposed in \cite{Bonanno:2023rzk}, with a particular emphasis on the behavior of higher overtone modes of QNMs. Utilizing both the Bernstein spectral method and the asymptotic iteration method---each well-suited for resolving higher overtones---we perform precise numerical calculations of the fundamental mode and several overtone frequencies.

Our analysis reveals that the fundamental mode exhibits only weak dependence on the parameter $\xi$. However, significant deviations arise in the QNM spectrum at higher overtones when comparing the regular BH to the Schwarzschild case. In particular, consistent with the findings reported in  \cite{Konoplya:2022hll, Konoplya:2022pbc, Konoplya:2022iyn, Gong:2023ghh}, we observe a pronounced outburst of higher overtones, suggesting that quantum corrections near the event horizon of the regular BH lead to substantial modifications in the QNM spectrum. This result offers novel insights into potential observational signatures and underlying physics near BH horizons. 

We further examine the correspondence between grey-body factors and QNMs proposed in Ref. \cite{Konoplya:2024lir} by employing the sixth-order WKB approximation to compute the grey-body factors, and then comparing them with those obtained by substituting accurate QNM frequencies into the correspondence. Our findings demonstrate high accuracy of this correspondence, with precision improving as the multipole number $\ell$ increases. Notably, the grey-body factors exhibit a gradual decrease with increasing $\xi$. This behavior is attributed to the enhancement of the effective potential, which reduces the transmission coefficient.

This study focuses exclusively on axial gravitational perturbations. In future work, it is interesting to extend our analysis to polar perturbations and explore high overtone modes in the context of rotating BHs.

\begin{acknowledgments}
	We are grateful to Antonio Bonanno, Huajie Gong, Daniele Malafarina and Qin Tan helpful discussions. The work is in part supported by NSFC Grant No.12205104 and the startup funding of South China University of Technology.
\end{acknowledgments}


\begin{thebibliography}{73}%
	\makeatletter
	\providecommand \@ifxundefined [1]{%
		\@ifx{#1\undefined}
	}%
	\providecommand \@ifnum [1]{%
		\ifnum #1\expandafter \@firstoftwo
		\else \expandafter \@secondoftwo
		\fi
	}%
	\providecommand \@ifx [1]{%
		\ifx #1\expandafter \@firstoftwo
		\else \expandafter \@secondoftwo
		\fi
	}%
	\providecommand \natexlab [1]{#1}%
	\providecommand \enquote  [1]{``#1''}%
	\providecommand \bibnamefont  [1]{#1}%
	\providecommand \bibfnamefont [1]{#1}%
	\providecommand \citenamefont [1]{#1}%
	\providecommand \href@noop [0]{\@secondoftwo}%
	\providecommand \href [0]{\begingroup \@sanitize@url \@href}%
	\providecommand \@href[1]{\@@startlink{#1}\@@href}%
	\providecommand \@@href[1]{\endgroup#1\@@endlink}%
	\providecommand \@sanitize@url [0]{\catcode `\\12\catcode `\$12\catcode
		`\&12\catcode `\#12\catcode `\^12\catcode `\_12\catcode `\%12\relax}%
	\providecommand \@@startlink[1]{}%
	\providecommand \@@endlink[0]{}%
	\providecommand \url  [0]{\begingroup\@sanitize@url \@url }%
	\providecommand \@url [1]{\endgroup\@href {#1}{\urlprefix }}%
	\providecommand \urlprefix  [0]{URL }%
	\providecommand \Eprint [0]{\href }%
	\providecommand \doibase [0]{http://dx.doi.org/}%
	\providecommand \selectlanguage [0]{\@gobble}%
	\providecommand \bibinfo  [0]{\@secondoftwo}%
	\providecommand \bibfield  [0]{\@secondoftwo}%
	\providecommand \translation [1]{[#1]}%
	\providecommand \BibitemOpen [0]{}%
	\providecommand \bibitemStop [0]{}%
	\providecommand \bibitemNoStop [0]{.\EOS\space}%
	\providecommand \EOS [0]{\spacefactor3000\relax}%
	\providecommand \BibitemShut  [1]{\csname bibitem#1\endcsname}%
	\let\auto@bib@innerbib\@empty
	\bibitem [{\citenamefont {Bonanno}\ \emph {et~al.}(2024)\citenamefont
		{Bonanno}, \citenamefont {Malafarina},\ and\ \citenamefont
		{Panassiti}}]{Bonanno:2023rzk}%
	\BibitemOpen
	\bibfield  {author} {\bibinfo {author} {\bibfnamefont {A.}~\bibnamefont
			{Bonanno}}, \bibinfo {author} {\bibfnamefont {D.}~\bibnamefont {Malafarina}},
		\ and\ \bibinfo {author} {\bibfnamefont {A.}~\bibnamefont {Panassiti}},\
	}\href {\doibase 10.1103/PhysRevLett.132.031401} {\bibfield  {journal}
		{\bibinfo  {journal} {Phys. Rev. Lett.}\ }\textbf {\bibinfo {volume} {132}},\
		\bibinfo {pages} {031401} (\bibinfo {year} {2024})},\ \Eprint
	{http://arxiv.org/abs/2308.10890} {arXiv:2308.10890 [gr-qc]} \BibitemShut
	{NoStop}%
	\bibitem [{\citenamefont {Penrose}(1965)}]{Penrose:1964wq}%
	\BibitemOpen
	\bibfield  {author} {\bibinfo {author} {\bibfnamefont {R.}~\bibnamefont
			{Penrose}},\ }\href {\doibase 10.1103/PhysRevLett.14.57} {\bibfield
		{journal} {\bibinfo  {journal} {Phys. Rev. Lett.}\ }\textbf {\bibinfo
			{volume} {14}},\ \bibinfo {pages} {57} (\bibinfo {year} {1965})}\BibitemShut
	{NoStop}%
	\bibitem [{\citenamefont {Hawking}\ and\ \citenamefont
		{Penrose}(1970)}]{Hawking:1970zqf}%
	\BibitemOpen
	\bibfield  {author} {\bibinfo {author} {\bibfnamefont {S.~W.}\ \bibnamefont
			{Hawking}}\ and\ \bibinfo {author} {\bibfnamefont {R.}~\bibnamefont
			{Penrose}},\ }\href {\doibase 10.1098/rspa.1970.0021} {\bibfield  {journal}
		{\bibinfo  {journal} {Proc. Roy. Soc. Lond. A}\ }\textbf {\bibinfo {volume}
			{314}},\ \bibinfo {pages} {529} (\bibinfo {year} {1970})}\BibitemShut
	{NoStop}%
	\bibitem [{\citenamefont {Bardeen}(1968)}]{Bardeen:1968bd}%
	\BibitemOpen
	\bibfield  {author} {\bibinfo {author} {\bibfnamefont {J.}~\bibnamefont
			{Bardeen}},\ }\href@noop {} {\bibfield  {journal} {\bibinfo  {journal}
			{Proceedings of the 5th International Conference on Gravitation and the
				Theory of Relativity}\ ,\ \bibinfo {pages} {87}} (\bibinfo {year}
		{1968})}\BibitemShut {NoStop}%
	\bibitem [{\citenamefont {Bambi}(2023)}]{Bambi:2023try}%
	\BibitemOpen
	\bibinfo {editor} {\bibfnamefont {C.}~\bibnamefont {Bambi}},\ ed.,\ \href
	{\doibase 10.1007/978-981-99-1596-5} {\emph {\bibinfo {title} {{Regular Black
					Holes. Towards a New Paradigm of Gravitational Collapse}}}},\ Springer Series
	in Astrophysics and Cosmology\ (\bibinfo  {publisher} {Springer},\ \bibinfo
	{year} {2023})\ \Eprint {http://arxiv.org/abs/2307.13249} {arXiv:2307.13249
		[gr-qc]} \BibitemShut {NoStop}%
	\bibitem [{\citenamefont {Platania}(2023)}]{Platania:2023srt}%
	\BibitemOpen
	\bibfield  {author} {\bibinfo {author} {\bibfnamefont {A.}~\bibnamefont
			{Platania}},\ }\href {\doibase 10.1007/978-981-19-3079-9_24-1} {\emph
		{\bibinfo {title} {{Black Holes in Asymptotically Safe Gravity}}}}\ (\bibinfo
	{year} {2023})\ \Eprint {http://arxiv.org/abs/2302.04272} {arXiv:2302.04272
		[gr-qc]} \BibitemShut {NoStop}%
         \bibitem [{\citenamefont {Spina}(2025)}]{Spina:2025wxb}%
    \BibitemOpen
    \bibfield  {author} {\bibinfo {author} {\bibfnamefont {A.}~\bibnamefont {Spina}},\ }\href {\doibase 10.53941/ijgtp.2025.100008} {\bibfield  {journal} {\bibinfo  {journal} {Int. J. Grav. Theor. Phys.}}\ \textbf {\bibinfo {volume} {1}},\ \bibinfo {number} {1},\ \bibinfo {pages} {8} (\bibinfo {year} {2025})},\ \Eprint {http://arxiv.org/abs/2510.14552} {arXiv:2510.14552 [gr-qc]} \BibitemShut {NoStop}%
	\bibitem [{\citenamefont {Bonanno}\ and\ \citenamefont
		{Reuter}(2000)}]{Bonanno:2000ep}%
	\BibitemOpen
	\bibfield  {author} {\bibinfo {author} {\bibfnamefont {A.}~\bibnamefont
			{Bonanno}}\ and\ \bibinfo {author} {\bibfnamefont {M.}~\bibnamefont
			{Reuter}},\ }\href {\doibase 10.1103/PhysRevD.62.043008} {\bibfield
		{journal} {\bibinfo  {journal} {Phys. Rev. D}\ }\textbf {\bibinfo {volume}
			{62}},\ \bibinfo {pages} {043008} (\bibinfo {year} {2000})},\ \Eprint
	{http://arxiv.org/abs/hep-th/0002196} {arXiv:hep-th/0002196} \BibitemShut
	{NoStop}%
	\bibitem [{\citenamefont {Markov}\ and\ \citenamefont
		{Mukhanov}(1985)}]{Markov:1985py}%
	\BibitemOpen
	\bibfield  {author} {\bibinfo {author} {\bibfnamefont {M.~A.}\ \bibnamefont
			{Markov}}\ and\ \bibinfo {author} {\bibfnamefont {V.~F.}\ \bibnamefont
			{Mukhanov}},\ }\href {\doibase 10.1007/BF02732276} {\bibfield  {journal}
		{\bibinfo  {journal} {Nuovo Cim. B}\ }\textbf {\bibinfo {volume} {86}},\
		\bibinfo {pages} {97} (\bibinfo {year} {1985})}\BibitemShut {NoStop}%
	\bibitem [{\citenamefont {Vishveshwara}(1970)}]{Vishveshwara:1970zz}%
	\BibitemOpen
	\bibfield  {author} {\bibinfo {author} {\bibfnamefont {C.~V.}\ \bibnamefont
			{Vishveshwara}},\ }\href {\doibase 10.1038/227936a0} {\bibfield  {journal}
		{\bibinfo  {journal} {Nature}\ }\textbf {\bibinfo {volume} {227}},\ \bibinfo
		{pages} {936} (\bibinfo {year} {1970})}\BibitemShut {NoStop}%
	\bibitem [{\citenamefont {Nollert}(1999)}]{Nollert:1999ji}%
	\BibitemOpen
	\bibfield  {author} {\bibinfo {author} {\bibfnamefont {H.-P.}\ \bibnamefont
			{Nollert}},\ }\href {\doibase 10.1088/0264-9381/16/12/201} {\bibfield
		{journal} {\bibinfo  {journal} {Class. Quant. Grav.}\ }\textbf {\bibinfo
			{volume} {16}},\ \bibinfo {pages} {R159} (\bibinfo {year}
		{1999})}\BibitemShut {NoStop}%
	\bibitem [{\citenamefont {Kokkotas}\ and\ \citenamefont
		{Schmidt}(1999)}]{Kokkotas:1999bd}%
	\BibitemOpen
	\bibfield  {author} {\bibinfo {author} {\bibfnamefont {K.~D.}\ \bibnamefont
			{Kokkotas}}\ and\ \bibinfo {author} {\bibfnamefont {B.~G.}\ \bibnamefont
			{Schmidt}},\ }\href {\doibase 10.12942/lrr-1999-2} {\bibfield  {journal}
		{\bibinfo  {journal} {Living Rev. Rel.}\ }\textbf {\bibinfo {volume} {2}},\
		\bibinfo {pages} {2} (\bibinfo {year} {1999})},\ \Eprint
	{http://arxiv.org/abs/gr-qc/9909058} {arXiv:gr-qc/9909058} \BibitemShut
	{NoStop}%
	\bibitem [{\citenamefont {Berti}\ \emph {et~al.}(2009)\citenamefont {Berti},
		\citenamefont {Cardoso},\ and\ \citenamefont {Starinets}}]{Berti:2009kk}%
	\BibitemOpen
	\bibfield  {author} {\bibinfo {author} {\bibfnamefont {E.}~\bibnamefont
			{Berti}}, \bibinfo {author} {\bibfnamefont {V.}~\bibnamefont {Cardoso}}, \
		and\ \bibinfo {author} {\bibfnamefont {A.~O.}\ \bibnamefont {Starinets}},\
	}\href {\doibase 10.1088/0264-9381/26/16/163001} {\bibfield  {journal}
		{\bibinfo  {journal} {Class. Quant. Grav.}\ }\textbf {\bibinfo {volume}
			{26}},\ \bibinfo {pages} {163001} (\bibinfo {year} {2009})},\ \Eprint
	{http://arxiv.org/abs/0905.2975} {arXiv:0905.2975 [gr-qc]} \BibitemShut
	{NoStop}%
	\bibitem [{\citenamefont {Konoplya}\ and\ \citenamefont
		{Zhidenko}(2011)}]{Konoplya:2011qq}%
	\BibitemOpen
	\bibfield  {author} {\bibinfo {author} {\bibfnamefont {R.~A.}\ \bibnamefont
			{Konoplya}}\ and\ \bibinfo {author} {\bibfnamefont {A.}~\bibnamefont
			{Zhidenko}},\ }\href {\doibase 10.1103/RevModPhys.83.793} {\bibfield
		{journal} {\bibinfo  {journal} {Rev. Mod. Phys.}\ }\textbf {\bibinfo {volume}
			{83}},\ \bibinfo {pages} {793} (\bibinfo {year} {2011})},\ \Eprint
	{http://arxiv.org/abs/1102.4014} {arXiv:1102.4014 [gr-qc]} \BibitemShut
	{NoStop}%
    \bibitem [{\citenamefont {Bolokhov}\ and\ \citenamefont {Skvortsova}(2025)}]{Bolokhov:2025rng}%
    \BibitemOpen
    \bibfield  {author} {\bibinfo {author} {\bibfnamefont {S.~V.}~\bibnamefont {Bolokhov}}\ and\ \bibinfo {author} {\bibfnamefont {M.}~\bibnamefont {Skvortsova}},\ }\href {\doibase 10.1134/S0202289325700306} {\bibfield  {journal} {\bibinfo  {journal} {Grav. Cosmol.}}\ \textbf {\bibinfo {volume} {31}},\ \bibinfo {number} {4},\ \bibinfo {pages} {423--446} (\bibinfo {year} {2025})} \BibitemShut {NoStop}%
	\bibitem [{\citenamefont {Berti}\ \emph {et~al.}(2018)\citenamefont {Berti},
		\citenamefont {Yagi}, \citenamefont {Yang},\ and\ \citenamefont
		{Yunes}}]{Berti:2018vdi}%
	\BibitemOpen
	\bibfield  {author} {\bibinfo {author} {\bibfnamefont {E.}~\bibnamefont
			{Berti}}, \bibinfo {author} {\bibfnamefont {K.}~\bibnamefont {Yagi}},
		\bibinfo {author} {\bibfnamefont {H.}~\bibnamefont {Yang}}, \ and\ \bibinfo
		{author} {\bibfnamefont {N.}~\bibnamefont {Yunes}},\ }\href {\doibase
		10.1007/s10714-018-2372-6} {\bibfield  {journal} {\bibinfo  {journal} {Gen.
				Rel. Grav.}\ }\textbf {\bibinfo {volume} {50}},\ \bibinfo {pages} {49}
		(\bibinfo {year} {2018})},\ \Eprint {http://arxiv.org/abs/1801.03587}
	{arXiv:1801.03587 [gr-qc]} \BibitemShut {NoStop}%
	\bibitem [{\citenamefont {Berti}\ \emph {et~al.}(2025)\citenamefont {Berti}
		\emph {et~al.}}]{Berti:2025hly}%
	\BibitemOpen
	\bibfield  {author} {\bibinfo {author} {\bibfnamefont {E.}~\bibnamefont
			{Berti}} \emph {et~al.},\ }\href@noop {} {\  (\bibinfo {year} {2025})},\
	\Eprint {http://arxiv.org/abs/2505.23895} {arXiv:2505.23895 [gr-qc]}
	\BibitemShut {NoStop}%
	\bibitem [{\citenamefont {Konoplya}\ and\ \citenamefont
		{Zhidenko}(2024{\natexlab{a}})}]{Konoplya:2022pbc}%
	\BibitemOpen
	\bibfield  {author} {\bibinfo {author} {\bibfnamefont {R.~A.}\ \bibnamefont
			{Konoplya}}\ and\ \bibinfo {author} {\bibfnamefont {A.}~\bibnamefont
			{Zhidenko}},\ }\href {\doibase 10.1016/j.jheap.2024.10.015} {\bibfield
		{journal} {\bibinfo  {journal} {JHEAp}\ }\textbf {\bibinfo {volume} {44}},\
		\bibinfo {pages} {419} (\bibinfo {year} {2024}{\natexlab{a}})},\ \Eprint
	{http://arxiv.org/abs/2209.00679} {arXiv:2209.00679 [gr-qc]} \BibitemShut
	{NoStop}%
	\bibitem [{\citenamefont {Giesler}\ \emph {et~al.}(2019)\citenamefont
		{Giesler}, \citenamefont {Isi}, \citenamefont {Scheel},\ and\ \citenamefont
		{Teukolsky}}]{Giesler:2019uxc}%
	\BibitemOpen
	\bibfield  {author} {\bibinfo {author} {\bibfnamefont {M.}~\bibnamefont
			{Giesler}}, \bibinfo {author} {\bibfnamefont {M.}~\bibnamefont {Isi}},
		\bibinfo {author} {\bibfnamefont {M.~A.}\ \bibnamefont {Scheel}}, \ and\
		\bibinfo {author} {\bibfnamefont {S.}~\bibnamefont {Teukolsky}},\ }\href
	{\doibase 10.1103/PhysRevX.9.041060} {\bibfield  {journal} {\bibinfo
			{journal} {Phys. Rev. X}\ }\textbf {\bibinfo {volume} {9}},\ \bibinfo {pages}
		{041060} (\bibinfo {year} {2019})},\ \Eprint
	{http://arxiv.org/abs/1903.08284} {arXiv:1903.08284 [gr-qc]} \BibitemShut
	{NoStop}%
	\bibitem [{\citenamefont {Giesler}\ \emph {et~al.}(2025)\citenamefont {Giesler}
		\emph {et~al.}}]{Giesler:2024hcr}%
	\BibitemOpen
	\bibfield  {author} {\bibinfo {author} {\bibfnamefont {M.}~\bibnamefont
			{Giesler}} \emph {et~al.},\ }\href {\doibase 10.1103/PhysRevD.111.084041}
	{\bibfield  {journal} {\bibinfo  {journal} {Phys. Rev. D}\ }\textbf {\bibinfo
			{volume} {111}},\ \bibinfo {pages} {084041} (\bibinfo {year} {2025})},\
	\Eprint {http://arxiv.org/abs/2411.11269} {arXiv:2411.11269 [gr-qc]}
	\BibitemShut {NoStop}%
	\bibitem [{\citenamefont {Baibhav}\ \emph {et~al.}(2023)\citenamefont
		{Baibhav}, \citenamefont {Cheung}, \citenamefont {Berti}, \citenamefont
		{Cardoso}, \citenamefont {Carullo}, \citenamefont {Cotesta}, \citenamefont
		{Del~Pozzo},\ and\ \citenamefont {Duque}}]{Baibhav:2023clw}%
	\BibitemOpen
	\bibfield  {author} {\bibinfo {author} {\bibfnamefont {V.}~\bibnamefont
			{Baibhav}}, \bibinfo {author} {\bibfnamefont {M.~H.-Y.}\ \bibnamefont
			{Cheung}}, \bibinfo {author} {\bibfnamefont {E.}~\bibnamefont {Berti}},
		\bibinfo {author} {\bibfnamefont {V.}~\bibnamefont {Cardoso}}, \bibinfo
		{author} {\bibfnamefont {G.}~\bibnamefont {Carullo}}, \bibinfo {author}
		{\bibfnamefont {R.}~\bibnamefont {Cotesta}}, \bibinfo {author} {\bibfnamefont
			{W.}~\bibnamefont {Del~Pozzo}}, \ and\ \bibinfo {author} {\bibfnamefont
			{F.}~\bibnamefont {Duque}},\ }\href {\doibase 10.1103/PhysRevD.108.104020}
	{\bibfield  {journal} {\bibinfo  {journal} {Phys. Rev. D}\ }\textbf {\bibinfo
			{volume} {108}},\ \bibinfo {pages} {104020} (\bibinfo {year} {2023})},\
	\Eprint {http://arxiv.org/abs/2302.03050} {arXiv:2302.03050 [gr-qc]}
	\BibitemShut {NoStop}%
	\bibitem [{\citenamefont {Konoplya}\ \emph {et~al.}(2022)\citenamefont
		{Konoplya}, \citenamefont {Zinhailo}, \citenamefont {Kunz}, \citenamefont
		{Stuchlik},\ and\ \citenamefont {Zhidenko}}]{Konoplya:2022hll}%
	\BibitemOpen
	\bibfield  {author} {\bibinfo {author} {\bibfnamefont {R.~A.}\ \bibnamefont
			{Konoplya}}, \bibinfo {author} {\bibfnamefont {A.~F.}\ \bibnamefont
			{Zinhailo}}, \bibinfo {author} {\bibfnamefont {J.}~\bibnamefont {Kunz}},
		\bibinfo {author} {\bibfnamefont {Z.}~\bibnamefont {Stuchlik}}, \ and\
		\bibinfo {author} {\bibfnamefont {A.}~\bibnamefont {Zhidenko}},\ }\href
	{\doibase 10.1088/1475-7516/2022/10/091} {\bibfield  {journal} {\bibinfo
			{journal} {JCAP}\ }\textbf {\bibinfo {volume} {10}},\ \bibinfo {pages} {091}
		(\bibinfo {year} {2022})},\ \Eprint {http://arxiv.org/abs/2206.14714}
	{arXiv:2206.14714 [gr-qc]} \BibitemShut {NoStop}%
	\bibitem [{\citenamefont {Konoplya}(2023{\natexlab{a}})}]{Konoplya:2022iyn}%
	\BibitemOpen
	\bibfield  {author} {\bibinfo {author} {\bibfnamefont {R.~A.}\ \bibnamefont
			{Konoplya}},\ }\href {\doibase 10.1103/PhysRevD.107.064039} {\bibfield
		{journal} {\bibinfo  {journal} {Phys. Rev. D}\ }\textbf {\bibinfo {volume}
			{107}},\ \bibinfo {pages} {064039} (\bibinfo {year} {2023}{\natexlab{a}})},\
	\Eprint {http://arxiv.org/abs/2210.14506} {arXiv:2210.14506 [gr-qc]}
	\BibitemShut {NoStop}%
	\bibitem [{\citenamefont {Fu}\ \emph {et~al.}(2024)\citenamefont {Fu},
		\citenamefont {Zhang}, \citenamefont {Liu}, \citenamefont {Kuang},\ and\
		\citenamefont {Wu}}]{Fu:2023drp}%
	\BibitemOpen
	\bibfield  {author} {\bibinfo {author} {\bibfnamefont {G.}~\bibnamefont
			{Fu}}, \bibinfo {author} {\bibfnamefont {D.}~\bibnamefont {Zhang}}, \bibinfo
		{author} {\bibfnamefont {P.}~\bibnamefont {Liu}}, \bibinfo {author}
		{\bibfnamefont {X.-M.}\ \bibnamefont {Kuang}}, \ and\ \bibinfo {author}
		{\bibfnamefont {J.-P.}\ \bibnamefont {Wu}},\ }\href {\doibase
		10.1103/PhysRevD.109.026010} {\bibfield  {journal} {\bibinfo  {journal}
			{Phys. Rev. D}\ }\textbf {\bibinfo {volume} {109}},\ \bibinfo {pages}
		{026010} (\bibinfo {year} {2024})},\ \Eprint
	{http://arxiv.org/abs/2301.08421} {arXiv:2301.08421 [gr-qc]} \BibitemShut
	{NoStop}%
	\bibitem [{\citenamefont {Moreira}\ \emph {et~al.}(2023)\citenamefont
		{Moreira}, \citenamefont {Lima~Junior}, \citenamefont {Crispino},\ and\
		\citenamefont {Herdeiro}}]{Moreira:2023cxy}%
	\BibitemOpen
	\bibfield  {author} {\bibinfo {author} {\bibfnamefont {Z.~S.}\ \bibnamefont
			{Moreira}}, \bibinfo {author} {\bibfnamefont {H.~C.~D.}\ \bibnamefont
			{Lima~Junior}}, \bibinfo {author} {\bibfnamefont {L.~C.~B.}\ \bibnamefont
			{Crispino}}, \ and\ \bibinfo {author} {\bibfnamefont {C.~A.~R.}\ \bibnamefont
			{Herdeiro}},\ }\href {\doibase 10.1103/PhysRevD.107.104016} {\bibfield
		{journal} {\bibinfo  {journal} {Phys. Rev. D}\ }\textbf {\bibinfo {volume}
			{107}},\ \bibinfo {pages} {104016} (\bibinfo {year} {2023})},\ \Eprint
	{http://arxiv.org/abs/2302.14722} {arXiv:2302.14722 [gr-qc]} \BibitemShut
	{NoStop}%
	\bibitem [{\citenamefont {Konoplya}\ \emph
		{et~al.}(2023{\natexlab{a}})\citenamefont {Konoplya}, \citenamefont
		{Stuchlik}, \citenamefont {Zhidenko},\ and\ \citenamefont
		{Zinhailo}}]{Konoplya:2023aph}%
	\BibitemOpen
	\bibfield  {author} {\bibinfo {author} {\bibfnamefont {R.~A.}\ \bibnamefont
			{Konoplya}}, \bibinfo {author} {\bibfnamefont {Z.}~\bibnamefont {Stuchlik}},
		\bibinfo {author} {\bibfnamefont {A.}~\bibnamefont {Zhidenko}}, \ and\
		\bibinfo {author} {\bibfnamefont {A.~F.}\ \bibnamefont {Zinhailo}},\ }\href
	{\doibase 10.1103/PhysRevD.107.104050} {\bibfield  {journal} {\bibinfo
			{journal} {Phys. Rev. D}\ }\textbf {\bibinfo {volume} {107}},\ \bibinfo
		{pages} {104050} (\bibinfo {year} {2023}{\natexlab{a}})},\ \Eprint
	{http://arxiv.org/abs/2303.01987} {arXiv:2303.01987 [gr-qc]} \BibitemShut
	{NoStop}%
	\bibitem [{\citenamefont {Konoplya}(2023{\natexlab{b}})}]{Konoplya:2023ppx}%
	\BibitemOpen
	\bibfield  {author} {\bibinfo {author} {\bibfnamefont {R.~A.}\ \bibnamefont
			{Konoplya}},\ }\href {\doibase 10.1088/1475-7516/2023/07/001} {\bibfield
		{journal} {\bibinfo  {journal} {JCAP}\ }\textbf {\bibinfo {volume} {07}},\
		\bibinfo {pages} {001} (\bibinfo {year} {2023}{\natexlab{b}})},\ \Eprint
	{http://arxiv.org/abs/2305.09187} {arXiv:2305.09187 [gr-qc]} \BibitemShut
	{NoStop}%
	\bibitem [{\citenamefont {Konoplya}\ \emph
		{et~al.}(2023{\natexlab{b}})\citenamefont {Konoplya}, \citenamefont
		{Ovchinnikov},\ and\ \citenamefont {Ahmedov}}]{Konoplya:2023ahd}%
	\BibitemOpen
	\bibfield  {author} {\bibinfo {author} {\bibfnamefont {R.~A.}\ \bibnamefont
			{Konoplya}}, \bibinfo {author} {\bibfnamefont {D.}~\bibnamefont
			{Ovchinnikov}}, \ and\ \bibinfo {author} {\bibfnamefont {B.}~\bibnamefont
			{Ahmedov}},\ }\href {\doibase 10.1103/PhysRevD.108.104054} {\bibfield
		{journal} {\bibinfo  {journal} {Phys. Rev. D}\ }\textbf {\bibinfo {volume}
			{108}},\ \bibinfo {pages} {104054} (\bibinfo {year} {2023}{\natexlab{b}})},\
	\Eprint {http://arxiv.org/abs/2307.10801} {arXiv:2307.10801 [gr-qc]}
	\BibitemShut {NoStop}%
	\bibitem [{\citenamefont {Bolokhov}(2024{\natexlab{a}})}]{Bolokhov:2023bwm}%
	\BibitemOpen
	\bibfield  {author} {\bibinfo {author} {\bibfnamefont {S.~V.}\ \bibnamefont
			{Bolokhov}},\ }\href {\doibase 10.1103/PhysRevD.110.024010} {\bibfield
		{journal} {\bibinfo  {journal} {Phys. Rev. D}\ }\textbf {\bibinfo {volume}
			{110}},\ \bibinfo {pages} {024010} (\bibinfo {year} {2024}{\natexlab{a}})},\
	\Eprint {http://arxiv.org/abs/2311.05503} {arXiv:2311.05503 [gr-qc]}
	\BibitemShut {NoStop}%
	\bibitem [{\citenamefont {Gong}\ \emph {et~al.}(2024)\citenamefont {Gong},
		\citenamefont {Li}, \citenamefont {Zhang}, \citenamefont {Fu},\ and\
		\citenamefont {Wu}}]{Gong:2023ghh}%
	\BibitemOpen
	\bibfield  {author} {\bibinfo {author} {\bibfnamefont {H.}~\bibnamefont
			{Gong}}, \bibinfo {author} {\bibfnamefont {S.}~\bibnamefont {Li}}, \bibinfo
		{author} {\bibfnamefont {D.}~\bibnamefont {Zhang}}, \bibinfo {author}
		{\bibfnamefont {G.}~\bibnamefont {Fu}}, \ and\ \bibinfo {author}
		{\bibfnamefont {J.-P.}\ \bibnamefont {Wu}},\ }\href {\doibase
		10.1103/PhysRevD.110.044040} {\bibfield  {journal} {\bibinfo  {journal}
			{Phys. Rev. D}\ }\textbf {\bibinfo {volume} {110}},\ \bibinfo {pages}
		{044040} (\bibinfo {year} {2024})},\ \Eprint
	{http://arxiv.org/abs/2312.17639} {arXiv:2312.17639 [gr-qc]} \BibitemShut
	{NoStop}%
	\bibitem [{\citenamefont {Cao}\ \emph {et~al.}(2024)\citenamefont {Cao},
		\citenamefont {Chen}, \citenamefont {Wu}, \citenamefont {Xie},\ and\
		\citenamefont {Zhou}}]{Cao:2024oud}%
	\BibitemOpen
	\bibfield  {author} {\bibinfo {author} {\bibfnamefont {L.-M.}\ \bibnamefont
			{Cao}}, \bibinfo {author} {\bibfnamefont {J.-N.}\ \bibnamefont {Chen}},
		\bibinfo {author} {\bibfnamefont {L.-B.}\ \bibnamefont {Wu}}, \bibinfo
		{author} {\bibfnamefont {L.}~\bibnamefont {Xie}}, \ and\ \bibinfo {author}
		{\bibfnamefont {Y.-S.}\ \bibnamefont {Zhou}},\ }\href {\doibase
		10.1007/s11433-024-2435-5} {\bibfield  {journal} {\bibinfo  {journal} {Sci.
				China Phys. Mech. Astron.}\ }\textbf {\bibinfo {volume} {67}},\ \bibinfo
		{pages} {100412} (\bibinfo {year} {2024})},\ \Eprint
	{http://arxiv.org/abs/2401.09907} {arXiv:2401.09907 [gr-qc]} \BibitemShut
	{NoStop}%
	\bibitem [{\citenamefont {Zhang}\ \emph {et~al.}(2024)\citenamefont {Zhang},
		\citenamefont {Gong}, \citenamefont {Fu}, \citenamefont {Wu},\ and\
		\citenamefont {Pan}}]{Zhang:2024nny}%
	\BibitemOpen
	\bibfield  {author} {\bibinfo {author} {\bibfnamefont {D.}~\bibnamefont
			{Zhang}}, \bibinfo {author} {\bibfnamefont {H.}~\bibnamefont {Gong}},
		\bibinfo {author} {\bibfnamefont {G.}~\bibnamefont {Fu}}, \bibinfo {author}
		{\bibfnamefont {J.-P.}\ \bibnamefont {Wu}}, \ and\ \bibinfo {author}
		{\bibfnamefont {Q.}~\bibnamefont {Pan}},\ }\href {\doibase
		10.1140/epjc/s10052-024-12928-x} {\bibfield  {journal} {\bibinfo  {journal}
			{Eur. Phys. J. C}\ }\textbf {\bibinfo {volume} {84}},\ \bibinfo {pages} {564}
		(\bibinfo {year} {2024})},\ \Eprint {http://arxiv.org/abs/2402.15085}
	{arXiv:2402.15085 [gr-qc]} \BibitemShut {NoStop}%
	\bibitem [{\citenamefont {Gingrich}(2024)}]{Gingrich:2024tuf}%
	\BibitemOpen
	\bibfield  {author} {\bibinfo {author} {\bibfnamefont {D.~M.}\ \bibnamefont
			{Gingrich}},\ }\href {\doibase 10.1103/PhysRevD.110.084045} {\bibfield
		{journal} {\bibinfo  {journal} {Phys. Rev. D}\ }\textbf {\bibinfo {volume}
			{110}},\ \bibinfo {pages} {084045} (\bibinfo {year} {2024})},\ \Eprint
	{http://arxiv.org/abs/2404.04447} {arXiv:2404.04447 [gr-qc]} \BibitemShut
	{NoStop}%
	\bibitem [{\citenamefont {Hirano}\ \emph {et~al.}(2024)\citenamefont {Hirano},
		\citenamefont {Kimura}, \citenamefont {Yamaguchi},\ and\ \citenamefont
		{Zhang}}]{Hirano:2024fgp}%
	\BibitemOpen
	\bibfield  {author} {\bibinfo {author} {\bibfnamefont {S.}~\bibnamefont
			{Hirano}}, \bibinfo {author} {\bibfnamefont {M.}~\bibnamefont {Kimura}},
		\bibinfo {author} {\bibfnamefont {M.}~\bibnamefont {Yamaguchi}}, \ and\
		\bibinfo {author} {\bibfnamefont {J.}~\bibnamefont {Zhang}},\ }\href
	{\doibase 10.1103/PhysRevD.110.024015} {\bibfield  {journal} {\bibinfo
			{journal} {Phys. Rev. D}\ }\textbf {\bibinfo {volume} {110}},\ \bibinfo
		{pages} {024015} (\bibinfo {year} {2024})},\ \Eprint
	{http://arxiv.org/abs/2404.09672} {arXiv:2404.09672 [gr-qc]} \BibitemShut
	{NoStop}%
	\bibitem [{\citenamefont {Dubinsky}(2024{\natexlab{a}})}]{Dubinsky:2024gwo}%
	\BibitemOpen
	\bibfield  {author} {\bibinfo {author} {\bibfnamefont {A.}~\bibnamefont
			{Dubinsky}},\ }\href {\doibase 10.1142/S0217732324501086} {\bibfield
		{journal} {\bibinfo  {journal} {Mod. Phys. Lett. A}\ }\textbf {\bibinfo
			{volume} {39}},\ \bibinfo {pages} {2450108} (\bibinfo {year}
		{2024}{\natexlab{a}})},\ \Eprint {http://arxiv.org/abs/2404.18004}
	{arXiv:2404.18004 [gr-qc]} \BibitemShut {NoStop}%
	\bibitem [{\citenamefont {Bolokhov}(2024{\natexlab{b}})}]{Bolokhov:2023ruj}%
	\BibitemOpen
	\bibfield  {author} {\bibinfo {author} {\bibfnamefont {S.~V.}\ \bibnamefont
			{Bolokhov}},\ }\href {\doibase 10.1103/PhysRevD.109.064017} {\bibfield
		{journal} {\bibinfo  {journal} {Phys. Rev. D}\ }\textbf {\bibinfo {volume}
			{109}},\ \bibinfo {pages} {064017} (\bibinfo {year}
		{2024}{\natexlab{b}})}\BibitemShut {NoStop}%
	\bibitem [{\citenamefont {Livine}\ \emph {et~al.}(2024)\citenamefont {Livine},
		\citenamefont {Montagnon}, \citenamefont {Oshita},\ and\ \citenamefont
		{Roussille}}]{Livine:2024bvo}%
	\BibitemOpen
	\bibfield  {author} {\bibinfo {author} {\bibfnamefont {E.~R.}\ \bibnamefont
			{Livine}}, \bibinfo {author} {\bibfnamefont {C.}~\bibnamefont {Montagnon}},
		\bibinfo {author} {\bibfnamefont {N.}~\bibnamefont {Oshita}}, \ and\ \bibinfo
		{author} {\bibfnamefont {H.}~\bibnamefont {Roussille}},\ }\href {\doibase
		10.1088/1475-7516/2024/10/037} {\bibfield  {journal} {\bibinfo  {journal}
			{JCAP}\ }\textbf {\bibinfo {volume} {10}},\ \bibinfo {pages} {037} (\bibinfo
		{year} {2024})},\ \Eprint {http://arxiv.org/abs/2405.12671} {arXiv:2405.12671
		[gr-qc]} \BibitemShut {NoStop}%
	\bibitem [{\citenamefont {Zhu}\ \emph {et~al.}(2025)\citenamefont {Zhu},
		\citenamefont {Fu}, \citenamefont {Li}, \citenamefont {Zhang},\ and\
		\citenamefont {Wu}}]{Zhu:2024wic}%
	\BibitemOpen
	\bibfield  {author} {\bibinfo {author} {\bibfnamefont {L.-G.}\ \bibnamefont
			{Zhu}}, \bibinfo {author} {\bibfnamefont {G.}~\bibnamefont {Fu}}, \bibinfo
		{author} {\bibfnamefont {S.}~\bibnamefont {Li}}, \bibinfo {author}
		{\bibfnamefont {D.}~\bibnamefont {Zhang}}, \ and\ \bibinfo {author}
		{\bibfnamefont {J.-P.}\ \bibnamefont {Wu}},\ }\href {\doibase
		10.1103/PhysRevD.111.104008} {\bibfield  {journal} {\bibinfo  {journal}
			{Phys. Rev. D}\ }\textbf {\bibinfo {volume} {111}},\ \bibinfo {pages}
		{104008} (\bibinfo {year} {2025})},\ \Eprint
	{http://arxiv.org/abs/2410.00543} {arXiv:2410.00543 [gr-qc]} \BibitemShut
	{NoStop}%
	\bibitem [{\citenamefont {Tang}\ \emph {et~al.}(2024)\citenamefont {Tang},
		\citenamefont {Ling}, \citenamefont {Jiang},\ and\ \citenamefont
		{Li}}]{Tang:2024txx}%
	\BibitemOpen
	\bibfield  {author} {\bibinfo {author} {\bibfnamefont {C.}~\bibnamefont
			{Tang}}, \bibinfo {author} {\bibfnamefont {Y.}~\bibnamefont {Ling}}, \bibinfo
		{author} {\bibfnamefont {Q.-Q.}\ \bibnamefont {Jiang}}, \ and\ \bibinfo
		{author} {\bibfnamefont {G.-P.}\ \bibnamefont {Li}},\ }\href {\doibase
		10.1140/epjc/s10052-024-13699-1} {\bibfield  {journal} {\bibinfo  {journal}
			{Eur. Phys. J. C}\ }\textbf {\bibinfo {volume} {84}},\ \bibinfo {pages}
		{1296} (\bibinfo {year} {2024})},\ \Eprint {http://arxiv.org/abs/2411.01764}
	{arXiv:2411.01764 [gr-qc]} \BibitemShut {NoStop}%
	\bibitem [{\citenamefont
		{L\"utf\"uo\u{g}lu}(2025{\natexlab{a}})}]{Lutfuoglu:2025ljm}%
	\BibitemOpen
	\bibfield  {author} {\bibinfo {author} {\bibfnamefont {B.~C.}\ \bibnamefont
			{L\"utf\"uo\u{g}lu}},\ }\href@noop {} {\  (\bibinfo {year}
		{2025}{\natexlab{a}})},\ \Eprint {http://arxiv.org/abs/2504.18482}
	{arXiv:2504.18482 [gr-qc]} \BibitemShut {NoStop}%
	\bibitem [{\citenamefont {Stashko}(2024)}]{Stashko:2024wuq}%
	\BibitemOpen
	\bibfield  {author} {\bibinfo {author} {\bibfnamefont {O.}~\bibnamefont
			{Stashko}},\ }\href {\doibase 10.1103/PhysRevD.110.084016} {\bibfield
		{journal} {\bibinfo  {journal} {Phys. Rev. D}\ }\textbf {\bibinfo {volume}
			{110}},\ \bibinfo {pages} {084016} (\bibinfo {year} {2024})},\ \Eprint
	{http://arxiv.org/abs/2407.07892} {arXiv:2407.07892 [gr-qc]} \BibitemShut
	{NoStop}%
	\bibitem [{\citenamefont {Spina}\ \emph {et~al.}(2024)\citenamefont {Spina},
		\citenamefont {Silveravalle},\ and\ \citenamefont {Bonanno}}]{Spina:2024npx}%
	\BibitemOpen
	\bibfield  {author} {\bibinfo {author} {\bibfnamefont {A.}~\bibnamefont
			{Spina}}, \bibinfo {author} {\bibfnamefont {S.}~\bibnamefont {Silveravalle}},
		\ and\ \bibinfo {author} {\bibfnamefont {A.}~\bibnamefont {Bonanno}},\ }in\
	\href@noop {} {\emph {\bibinfo {booktitle} {{17th Marcel Grossmann Meeting}:
				{On Recent Developments in Theoretical and Experimental General Relativity,
					Gravitation, and Relativistic Field Theories}}}}\ (\bibinfo {year} {2024})\
	\Eprint {http://arxiv.org/abs/2410.05936} {arXiv:2410.05936 [gr-qc]}
	\BibitemShut {NoStop}%
	\bibitem [{\citenamefont {S\'anchez}(2024)}]{Sanchez:2024sdm}%
	\BibitemOpen
	\bibfield  {author} {\bibinfo {author} {\bibfnamefont {L.~A.}\ \bibnamefont
			{S\'anchez}},\ }\href {\doibase 10.1140/epjc/s10052-024-13398-x} {\bibfield
		{journal} {\bibinfo  {journal} {Eur. Phys. J. C}\ }\textbf {\bibinfo {volume}
			{84}},\ \bibinfo {pages} {1056} (\bibinfo {year} {2024})},\ \Eprint
	{http://arxiv.org/abs/2408.00226} {arXiv:2408.00226 [gr-qc]} \BibitemShut
	{NoStop}%
    \bibitem [{\citenamefont
		{L\"utf\"uo\u{g}lu}(2025{\natexlab{c}})}]{Lutfuoglu:2025ohb}%
	\BibitemOpen
	\bibfield  {author} {\bibinfo {author} {\bibfnamefont {B.~C.}\ \bibnamefont
			{L\"utf\"uo\u{g}lu}},\ }\href@noop {} {\  (\bibinfo {year}
		{2025}{\natexlab{c}})},\ \Eprint {http://arxiv.org/abs/2505.06966}
	{arXiv:2505.06966 [gr-qc]} \BibitemShut {NoStop}%
	\bibitem [{\citenamefont {Konoplya}\ and\ \citenamefont
		{Zhidenko}(2024{\natexlab{b}})}]{Konoplya:2024lir}%
	\BibitemOpen
	\bibfield  {author} {\bibinfo {author} {\bibfnamefont {R.~A.}\ \bibnamefont
			{Konoplya}}\ and\ \bibinfo {author} {\bibfnamefont {A.}~\bibnamefont
			{Zhidenko}},\ }\href {\doibase 10.1088/1475-7516/2024/09/068} {\bibfield
		{journal} {\bibinfo  {journal} {JCAP}\ }\textbf {\bibinfo {volume} {09}},\
		\bibinfo {pages} {068} (\bibinfo {year} {2024}{\natexlab{b}})},\ \Eprint
	{http://arxiv.org/abs/2406.11694} {arXiv:2406.11694 [gr-qc]} \BibitemShut
	{NoStop}%
        \bibitem [{\citenamefont {Bonanno}\ \emph{et~al.}(2025)}]{Bonanno:2025dry}%
    \BibitemOpen
    \bibfield  {author} {\bibinfo {author} {\bibfnamefont {A.~M.}~\bibnamefont {Bonanno}}\ and\ \bibinfo {author} {\bibfnamefont {R.~A.}~\bibnamefont {Konoplya}}\ and\ \bibinfo {author} {\bibfnamefont {G.}~\bibnamefont {Oglialoro}}\ and\ \bibinfo {author} {\bibfnamefont {A.}~\bibnamefont {Spina}},\ }\href {\doibase 10.1088/1475-7516/2025/12/042} {\bibfield  {journal} {\bibinfo  {journal} {JCAP}}\ \textbf {\bibinfo {volume} {12}},\ \bibinfo {pages} {042} (\bibinfo {year} {2025})},\ \Eprint {http://arxiv.org/abs/2509.12469} {arXiv:2509.12469 [gr-qc]} \BibitemShut {NoStop}%
	\bibitem [{\citenamefont {Konoplya}\ and\ \citenamefont
		{Zhidenko}(2025)}]{Konoplya:2024vuj}%
	\BibitemOpen
	\bibfield  {author} {\bibinfo {author} {\bibfnamefont {R.~A.}\ \bibnamefont
			{Konoplya}}\ and\ \bibinfo {author} {\bibfnamefont {A.}~\bibnamefont
			{Zhidenko}},\ }\href {\doibase 10.1016/j.physletb.2025.139288} {\bibfield
		{journal} {\bibinfo  {journal} {Phys. Lett. B}\ }\textbf {\bibinfo {volume}
			{861}},\ \bibinfo {pages} {139288} (\bibinfo {year} {2025})},\ \Eprint
	{http://arxiv.org/abs/2408.11162} {arXiv:2408.11162 [gr-qc]} \BibitemShut
	{NoStop}%
	\bibitem [{\citenamefont {Oshita}(2024)}]{Oshita:2023cjz}%
	\BibitemOpen
	\bibfield  {author} {\bibinfo {author} {\bibfnamefont {N.}~\bibnamefont
			{Oshita}},\ }\href {\doibase 10.1103/PhysRevD.109.104028} {\bibfield
		{journal} {\bibinfo  {journal} {Phys. Rev. D}\ }\textbf {\bibinfo {volume}
			{109}},\ \bibinfo {pages} {104028} (\bibinfo {year} {2024})},\ \Eprint
	{http://arxiv.org/abs/2309.05725} {arXiv:2309.05725 [gr-qc]} \BibitemShut
	{NoStop}%
	\bibitem [{\citenamefont {Okabayashi}\ and\ \citenamefont
		{Oshita}(2024)}]{Okabayashi:2024qbz}%
	\BibitemOpen
	\bibfield  {author} {\bibinfo {author} {\bibfnamefont {K.}~\bibnamefont
			{Okabayashi}}\ and\ \bibinfo {author} {\bibfnamefont {N.}~\bibnamefont
			{Oshita}},\ }\href {\doibase 10.1103/PhysRevD.110.064086} {\bibfield
		{journal} {\bibinfo  {journal} {Phys. Rev. D}\ }\textbf {\bibinfo {volume}
			{110}},\ \bibinfo {pages} {064086} (\bibinfo {year} {2024})},\ \Eprint
	{http://arxiv.org/abs/2403.17487} {arXiv:2403.17487 [gr-qc]} \BibitemShut
	{NoStop}%
	\bibitem [{\citenamefont {Rosato}\ \emph {et~al.}(2024)\citenamefont {Rosato},
		\citenamefont {Destounis},\ and\ \citenamefont {Pani}}]{Rosato:2024arw}%
	\BibitemOpen
	\bibfield  {author} {\bibinfo {author} {\bibfnamefont {R.~F.}\ \bibnamefont
			{Rosato}}, \bibinfo {author} {\bibfnamefont {K.}~\bibnamefont {Destounis}}, \
		and\ \bibinfo {author} {\bibfnamefont {P.}~\bibnamefont {Pani}},\ }\href
	{\doibase 10.1103/PhysRevD.110.L121501} {\bibfield  {journal} {\bibinfo
			{journal} {Phys. Rev. D}\ }\textbf {\bibinfo {volume} {110}},\ \bibinfo
		{pages} {L121501} (\bibinfo {year} {2024})},\ \Eprint
	{http://arxiv.org/abs/2406.01692} {arXiv:2406.01692 [gr-qc]} \BibitemShut
	{NoStop}%
	\bibitem [{\citenamefont {Oshita}\ \emph {et~al.}(2024)\citenamefont {Oshita},
		\citenamefont {Takahashi},\ and\ \citenamefont {Mukohyama}}]{Oshita:2024fzf}%
	\BibitemOpen
	\bibfield  {author} {\bibinfo {author} {\bibfnamefont {N.}~\bibnamefont
			{Oshita}}, \bibinfo {author} {\bibfnamefont {K.}~\bibnamefont {Takahashi}}, \
		and\ \bibinfo {author} {\bibfnamefont {S.}~\bibnamefont {Mukohyama}},\ }\href
	{\doibase 10.1103/PhysRevD.110.084070} {\bibfield  {journal} {\bibinfo
			{journal} {Phys. Rev. D}\ }\textbf {\bibinfo {volume} {110}},\ \bibinfo
		{pages} {084070} (\bibinfo {year} {2024})},\ \Eprint
	{http://arxiv.org/abs/2406.04525} {arXiv:2406.04525 [gr-qc]} \BibitemShut
	{NoStop}%
    \bibitem [{\citenamefont {Konoplya}\ and\ \citenamefont {Pappas}(2025)}]{Konoplya:2025ixm}%
    \BibitemOpen
    \bibfield  {author} {\bibinfo {author} {\bibfnamefont {R.~A.}~\bibnamefont {Konoplya}}\ and\ \bibinfo {author} {\bibfnamefont {T.~D.}~\bibnamefont {Pappas}},\ }\href@noop {} {\  (\bibinfo {year} {2025})},\ \Eprint {http://arxiv.org/abs/2507.01954} {arXiv:2507.01954 [gr-qc]} \BibitemShut {NoStop}%
	\bibitem [{\citenamefont {Skvortsova}(2024)}]{Skvortsova:2024msa}%
	\BibitemOpen
	\bibfield  {author} {\bibinfo {author} {\bibfnamefont {M.}~\bibnamefont
			{Skvortsova}},\ }\href@noop {} {\  (\bibinfo {year} {2024})},\ \Eprint
	{http://arxiv.org/abs/2411.06007} {arXiv:2411.06007 [gr-qc]} \BibitemShut
	{NoStop}%
	\bibitem [{\citenamefont {Dubinsky}(2024{\natexlab{b}})}]{Dubinsky:2024vbn}%
	\BibitemOpen
	\bibfield  {author} {\bibinfo {author} {\bibfnamefont {A.}~\bibnamefont
			{Dubinsky}},\ }\href@noop {} {\  (\bibinfo {year} {2024}{\natexlab{b}})},\
	\Eprint {http://arxiv.org/abs/2412.00625} {arXiv:2412.00625 [gr-qc]}
	\BibitemShut {NoStop}%
	\bibitem [{\citenamefont {Bolokhov}\ and\ \citenamefont
		{Skvortsova}(2025)}]{Bolokhov:2024otn}%
	\BibitemOpen
	\bibfield  {author} {\bibinfo {author} {\bibfnamefont {S.~V.}\ \bibnamefont
			{Bolokhov}}\ and\ \bibinfo {author} {\bibfnamefont {M.}~\bibnamefont
			{Skvortsova}},\ }\href {\doibase 10.1088/1475-7516/2025/04/025} {\bibfield
		{journal} {\bibinfo  {journal} {JCAP}\ }\textbf {\bibinfo {volume} {04}},\
		\bibinfo {pages} {025} (\bibinfo {year} {2025})},\ \Eprint
	{http://arxiv.org/abs/2412.11166} {arXiv:2412.11166 [gr-qc]} \BibitemShut
	{NoStop}%
	\bibitem [{\citenamefont {Malik}(2024)}]{Malik:2024wvs}%
	\BibitemOpen
	\bibfield  {author} {\bibinfo {author} {\bibfnamefont {Z.}~\bibnamefont
			{Malik}},\ }\href@noop {} {\  (\bibinfo {year} {2024})},\ \Eprint
	{http://arxiv.org/abs/2412.13385} {arXiv:2412.13385 [gr-qc]} \BibitemShut
	{NoStop}%
	\bibitem [{\citenamefont {Malik}(2025)}]{Malik:2024cgb}%
	\BibitemOpen
	\bibfield  {author} {\bibinfo {author} {\bibfnamefont {Z.}~\bibnamefont
			{Malik}},\ }\href {\doibase 10.1088/1475-7516/2025/04/042} {\bibfield
		{journal} {\bibinfo  {journal} {JCAP}\ }\textbf {\bibinfo {volume} {04}},\
		\bibinfo {pages} {042} (\bibinfo {year} {2025})},\ \Eprint
	{http://arxiv.org/abs/2412.19443} {arXiv:2412.19443 [gr-qc]} \BibitemShut
	{NoStop}%
	\bibitem [{\citenamefont
		{L\"utf\"uo\u{g}lu}(2025{\natexlab{b}})}]{Lutfuoglu:2025hjy}%
	\BibitemOpen
	\bibfield  {author} {\bibinfo {author} {\bibfnamefont {B.~C.}\ \bibnamefont
			{L\"utf\"uo\u{g}lu}},\ }\href {\doibase 10.1140/epjc/s10052-025-14210-0}
	{\bibfield  {journal} {\bibinfo  {journal} {Eur. Phys. J. C}\ }\textbf
		{\bibinfo {volume} {85}},\ \bibinfo {pages} {486} (\bibinfo {year}
		{2025}{\natexlab{b}})},\ \Eprint {http://arxiv.org/abs/2503.16087}
	{arXiv:2503.16087 [gr-qc]} \BibitemShut {NoStop}%
	\bibitem [{\citenamefont {Hamil}\ and\ \citenamefont
		{L\"utf\"uo\u{g}lu}(2025)}]{Hamil:2025cms}%
	\BibitemOpen
	\bibfield  {author} {\bibinfo {author} {\bibfnamefont {B.}~\bibnamefont
			{Hamil}}\ and\ \bibinfo {author} {\bibfnamefont {B.~C.}\ \bibnamefont
			{L\"utf\"uo\u{g}lu}},\ }\href@noop {} {\  (\bibinfo {year} {2025})},\ \Eprint
	{http://arxiv.org/abs/2503.17474} {arXiv:2503.17474 [gr-qc]} \BibitemShut
	{NoStop}%
	\bibitem [{\citenamefont {Tang}\ \emph {et~al.}(2025)\citenamefont {Tang},
		\citenamefont {Ling},\ and\ \citenamefont {Jiang}}]{Tang:2025mkk}%
	\BibitemOpen
	\bibfield  {author} {\bibinfo {author} {\bibfnamefont {C.}~\bibnamefont
			{Tang}}, \bibinfo {author} {\bibfnamefont {Y.}~\bibnamefont {Ling}}, \ and\
		\bibinfo {author} {\bibfnamefont {Q.-Q.}\ \bibnamefont {Jiang}},\ }\href@noop
	{} {\  (\bibinfo {year} {2025})},\ \Eprint {http://arxiv.org/abs/2503.21597}
	{arXiv:2503.21597 [gr-qc]} \BibitemShut {NoStop}%
	\bibitem [{\citenamefont {Pedrotti}\ and\ \citenamefont
		{Calz\`a}(2025)}]{Pedrotti:2025upg}%
	\BibitemOpen
	\bibfield  {author} {\bibinfo {author} {\bibfnamefont {D.}~\bibnamefont
			{Pedrotti}}\ and\ \bibinfo {author} {\bibfnamefont {M.}~\bibnamefont
			{Calz\`a}},\ }\href@noop {} {\  (\bibinfo {year} {2025})},\ \Eprint
	{http://arxiv.org/abs/2504.01909} {arXiv:2504.01909 [gr-qc]} \BibitemShut
	{NoStop}%
	\bibitem [{\citenamefont {Liang}\ \emph {et~al.}(2025)\citenamefont {Liang},
		\citenamefont {Liu},\ and\ \citenamefont {Long}}]{Liang:2025jph}%
	\BibitemOpen
	\bibfield  {author} {\bibinfo {author} {\bibfnamefont {J.}~\bibnamefont
			{Liang}}, \bibinfo {author} {\bibfnamefont {D.}~\bibnamefont {Liu}}, \ and\
		\bibinfo {author} {\bibfnamefont {Z.-W.}\ \bibnamefont {Long}},\ }\href@noop
	{} {\  (\bibinfo {year} {2025})},\ \Eprint {http://arxiv.org/abs/2504.12743}
	{arXiv:2504.12743 [gr-qc]} \BibitemShut {NoStop}%
	\bibitem [{\citenamefont {Xie}\ \emph {et~al.}(2025)\citenamefont {Xie},
		\citenamefont {Wu},\ and\ \citenamefont {Guo}}]{Xie:2025jbr}%
	\BibitemOpen
	\bibfield  {author} {\bibinfo {author} {\bibfnamefont {L.}~\bibnamefont
			{Xie}}, \bibinfo {author} {\bibfnamefont {L.-B.}\ \bibnamefont {Wu}}, \ and\
		\bibinfo {author} {\bibfnamefont {Z.-K.}\ \bibnamefont {Guo}},\ }\href@noop
	{} {\  (\bibinfo {year} {2025})},\ \Eprint {http://arxiv.org/abs/2505.21303}
	{arXiv:2505.21303 [gr-qc]} \BibitemShut {NoStop}%
	\bibitem [{\citenamefont {Konoplya}(2003)}]{Konoplya:2003ii}%
	\BibitemOpen
	\bibfield  {author} {\bibinfo {author} {\bibfnamefont {R.~A.}\ \bibnamefont
			{Konoplya}},\ }\href {\doibase 10.1103/PhysRevD.68.024018} {\bibfield
		{journal} {\bibinfo  {journal} {Phys. Rev. D}\ }\textbf {\bibinfo {volume}
			{68}},\ \bibinfo {pages} {024018} (\bibinfo {year} {2003})},\ \Eprint
	{http://arxiv.org/abs/gr-qc/0303052} {arXiv:gr-qc/0303052} \BibitemShut
	{NoStop}%
	\bibitem [{\citenamefont {Toshmatov}\ \emph {et~al.}(2017)\citenamefont
		{Toshmatov}, \citenamefont {Bambi}, \citenamefont {Ahmedov}, \citenamefont
		{Abdujabbarov},\ and\ \citenamefont {Stuchl\'\i{}k}}]{Toshmatov:2017kmw}%
	\BibitemOpen
	\bibfield  {author} {\bibinfo {author} {\bibfnamefont {B.}~\bibnamefont
			{Toshmatov}}, \bibinfo {author} {\bibfnamefont {C.}~\bibnamefont {Bambi}},
		\bibinfo {author} {\bibfnamefont {B.}~\bibnamefont {Ahmedov}}, \bibinfo
		{author} {\bibfnamefont {A.}~\bibnamefont {Abdujabbarov}}, \ and\ \bibinfo
		{author} {\bibfnamefont {Z.}~\bibnamefont {Stuchl\'\i{}k}},\ }\href {\doibase
		10.1140/epjc/s10052-017-5112-2} {\bibfield  {journal} {\bibinfo  {journal}
			{Eur. Phys. J. C}\ }\textbf {\bibinfo {volume} {77}},\ \bibinfo {pages} {542}
		(\bibinfo {year} {2017})},\ \Eprint {http://arxiv.org/abs/1702.06855}
	{arXiv:1702.06855 [gr-qc]} \BibitemShut {NoStop}%
	\bibitem [{\citenamefont {Ashtekar}\ \emph {et~al.}(2018)\citenamefont
		{Ashtekar}, \citenamefont {Olmedo},\ and\ \citenamefont
		{Singh}}]{Ashtekar:2018cay}%
	\BibitemOpen
	\bibfield  {author} {\bibinfo {author} {\bibfnamefont {A.}~\bibnamefont
			{Ashtekar}}, \bibinfo {author} {\bibfnamefont {J.}~\bibnamefont {Olmedo}}, \
		and\ \bibinfo {author} {\bibfnamefont {P.}~\bibnamefont {Singh}},\ }\href
	{\doibase 10.1103/PhysRevD.98.126003} {\bibfield  {journal} {\bibinfo
			{journal} {Phys. Rev. D}\ }\textbf {\bibinfo {volume} {98}},\ \bibinfo
		{pages} {126003} (\bibinfo {year} {2018})},\ \Eprint
	{http://arxiv.org/abs/1806.02406} {arXiv:1806.02406 [gr-qc]} \BibitemShut
	{NoStop}%
	\bibitem [{\citenamefont {Chen}\ and\ \citenamefont
		{Chen}(2019)}]{Chen:2019iuo}%
	\BibitemOpen
	\bibfield  {author} {\bibinfo {author} {\bibfnamefont {C.-Y.}\ \bibnamefont
			{Chen}}\ and\ \bibinfo {author} {\bibfnamefont {P.}~\bibnamefont {Chen}},\
	}\href {\doibase 10.1103/PhysRevD.99.104003} {\bibfield  {journal} {\bibinfo
			{journal} {Phys. Rev. D}\ }\textbf {\bibinfo {volume} {99}},\ \bibinfo
		{pages} {104003} (\bibinfo {year} {2019})},\ \Eprint
	{http://arxiv.org/abs/1902.01678} {arXiv:1902.01678 [gr-qc]} \BibitemShut
	{NoStop}%
	\bibitem [{\citenamefont {Yang}\ \emph {et~al.}(2023)\citenamefont {Yang},
		\citenamefont {Guo}, \citenamefont {Tan},\ and\ \citenamefont
		{Liu}}]{Yang:2023gas}%
	\BibitemOpen
	\bibfield  {author} {\bibinfo {author} {\bibfnamefont {S.}~\bibnamefont
			{Yang}}, \bibinfo {author} {\bibfnamefont {W.-D.}\ \bibnamefont {Guo}},
		\bibinfo {author} {\bibfnamefont {Q.}~\bibnamefont {Tan}}, \ and\ \bibinfo
		{author} {\bibfnamefont {Y.-X.}\ \bibnamefont {Liu}},\ }\href {\doibase
		10.1103/PhysRevD.108.024055} {\bibfield  {journal} {\bibinfo  {journal}
			{Phys. Rev. D}\ }\textbf {\bibinfo {volume} {108}},\ \bibinfo {pages}
		{024055} (\bibinfo {year} {2023})},\ \Eprint
	{http://arxiv.org/abs/2304.06895} {arXiv:2304.06895 [gr-qc]} \BibitemShut
	{NoStop}%
	\bibitem [{\citenamefont {Konoplya}\ and\ \citenamefont
		{Stashko}(2024)}]{Konoplya:2024lch}%
	\BibitemOpen
	\bibfield  {author} {\bibinfo {author} {\bibfnamefont {R.~A.}\ \bibnamefont
			{Konoplya}}\ and\ \bibinfo {author} {\bibfnamefont {O.~S.}\ \bibnamefont
			{Stashko}},\ }\href@noop {} {\  (\bibinfo {year} {2024})},\ \Eprint
	{http://arxiv.org/abs/2408.02578} {arXiv:2408.02578 [gr-qc]} \BibitemShut
	{NoStop}%
	\bibitem [{\citenamefont {Regge}\ and\ \citenamefont
		{Wheeler}(1957)}]{Regge:1957td}%
	\BibitemOpen
	\bibfield  {author} {\bibinfo {author} {\bibfnamefont {T.}~\bibnamefont
			{Regge}}\ and\ \bibinfo {author} {\bibfnamefont {J.~A.}\ \bibnamefont
			{Wheeler}},\ }\href {\doibase 10.1103/PhysRev.108.1063} {\bibfield  {journal}
		{\bibinfo  {journal} {Phys. Rev.}\ }\textbf {\bibinfo {volume} {108}},\
		\bibinfo {pages} {1063} (\bibinfo {year} {1957})}\BibitemShut {NoStop}%
	\bibitem [{\citenamefont {Cardoso}\ \emph {et~al.}(2022)\citenamefont
		{Cardoso}, \citenamefont {Destounis}, \citenamefont {Duque}, \citenamefont
		{Panosso~Macedo},\ and\ \citenamefont {Maselli}}]{Cardoso:2022whc}%
	\BibitemOpen
	\bibfield  {author} {\bibinfo {author} {\bibfnamefont {V.}~\bibnamefont
			{Cardoso}}, \bibinfo {author} {\bibfnamefont {K.}~\bibnamefont {Destounis}},
		\bibinfo {author} {\bibfnamefont {F.}~\bibnamefont {Duque}}, \bibinfo
		{author} {\bibfnamefont {R.}~\bibnamefont {Panosso~Macedo}}, \ and\ \bibinfo
		{author} {\bibfnamefont {A.}~\bibnamefont {Maselli}},\ }\href {\doibase
		10.1103/PhysRevLett.129.241103} {\bibfield  {journal} {\bibinfo  {journal}
			{Phys. Rev. Lett.}\ }\textbf {\bibinfo {volume} {129}},\ \bibinfo {pages}
		{241103} (\bibinfo {year} {2022})},\ \Eprint
	{http://arxiv.org/abs/2210.01133} {arXiv:2210.01133 [gr-qc]} \BibitemShut
	{NoStop}%
	\bibitem [{\citenamefont {Fortuna}\ and\ \citenamefont
		{Vega}(2023)}]{Fortuna:2020obg}%
	\BibitemOpen
	\bibfield  {author} {\bibinfo {author} {\bibfnamefont {S.}~\bibnamefont
			{Fortuna}}\ and\ \bibinfo {author} {\bibfnamefont {I.}~\bibnamefont {Vega}},\
	}\href {\doibase 10.1140/epjc/s10052-023-12350-9} {\bibfield  {journal}
		{\bibinfo  {journal} {Eur. Phys. J. C}\ }\textbf {\bibinfo {volume} {83}},\
		\bibinfo {pages} {1170} (\bibinfo {year} {2023})},\ \Eprint
	{http://arxiv.org/abs/2003.06232} {arXiv:2003.06232 [gr-qc]} \BibitemShut
	{NoStop}%
	\bibitem [{\citenamefont {Cho}\ \emph {et~al.}(2010)\citenamefont {Cho},
		\citenamefont {Cornell}, \citenamefont {Doukas},\ and\ \citenamefont
		{Naylor}}]{Cho:2009cj}%
	\BibitemOpen
	\bibfield  {author} {\bibinfo {author} {\bibfnamefont {H.~T.}\ \bibnamefont
			{Cho}}, \bibinfo {author} {\bibfnamefont {A.~S.}\ \bibnamefont {Cornell}},
		\bibinfo {author} {\bibfnamefont {J.}~\bibnamefont {Doukas}}, \ and\ \bibinfo
		{author} {\bibfnamefont {W.}~\bibnamefont {Naylor}},\ }\href {\doibase
		10.1088/0264-9381/27/15/155004} {\bibfield  {journal} {\bibinfo  {journal}
			{Class. Quant. Grav.}\ }\textbf {\bibinfo {volume} {27}},\ \bibinfo {pages}
		{155004} (\bibinfo {year} {2010})},\ \Eprint {http://arxiv.org/abs/0912.2740}
	{arXiv:0912.2740 [gr-qc]} \BibitemShut {NoStop}%
	\bibitem [{\citenamefont {Cho}\ \emph {et~al.}(2012)\citenamefont {Cho},
		\citenamefont {Cornell}, \citenamefont {Doukas}, \citenamefont {Huang},\ and\
		\citenamefont {Naylor}}]{Cho:2011sf}%
	\BibitemOpen
	\bibfield  {author} {\bibinfo {author} {\bibfnamefont {H.~T.}\ \bibnamefont
			{Cho}}, \bibinfo {author} {\bibfnamefont {A.~S.}\ \bibnamefont {Cornell}},
		\bibinfo {author} {\bibfnamefont {J.}~\bibnamefont {Doukas}}, \bibinfo
		{author} {\bibfnamefont {T.~R.}\ \bibnamefont {Huang}}, \ and\ \bibinfo
		{author} {\bibfnamefont {W.}~\bibnamefont {Naylor}},\ }\href {\doibase
		10.1155/2012/281705} {\bibfield  {journal} {\bibinfo  {journal} {Adv. Math.
				Phys.}\ }\textbf {\bibinfo {volume} {2012}},\ \bibinfo {pages} {281705}
		(\bibinfo {year} {2012})},\ \Eprint {http://arxiv.org/abs/1111.5024}
	{arXiv:1111.5024 [gr-qc]} \BibitemShut {NoStop}%
	\bibitem [{\citenamefont {Iyer}\ and\ \citenamefont
		{Will}(1987)}]{Iyer:1986np}%
	\BibitemOpen
	\bibfield  {author} {\bibinfo {author} {\bibfnamefont {S.}~\bibnamefont
			{Iyer}}\ and\ \bibinfo {author} {\bibfnamefont {C.~M.}\ \bibnamefont
			{Will}},\ }\href {\doibase 10.1103/PhysRevD.35.3621} {\bibfield  {journal}
		{\bibinfo  {journal} {Phys. Rev. D}\ }\textbf {\bibinfo {volume} {35}},\
		\bibinfo {pages} {3621} (\bibinfo {year} {1987})}\BibitemShut {NoStop}%
	\bibitem [{\citenamefont {Konoplya}\ \emph {et~al.}(2019)\citenamefont
		{Konoplya}, \citenamefont {Zhidenko},\ and\ \citenamefont
		{Zinhailo}}]{Konoplya:2019hlu}%
	\BibitemOpen
	\bibfield  {author} {\bibinfo {author} {\bibfnamefont {R.~A.}\ \bibnamefont
			{Konoplya}}, \bibinfo {author} {\bibfnamefont {A.}~\bibnamefont {Zhidenko}},
		\ and\ \bibinfo {author} {\bibfnamefont {A.~F.}\ \bibnamefont {Zinhailo}},\
	}\href {\doibase 10.1088/1361-6382/ab2e25} {\bibfield  {journal} {\bibinfo
			{journal} {Class. Quant. Grav.}\ }\textbf {\bibinfo {volume} {36}},\ \bibinfo
		{pages} {155002} (\bibinfo {year} {2019})},\ \Eprint
	{http://arxiv.org/abs/1904.10333} {arXiv:1904.10333 [gr-qc]} \BibitemShut
	{NoStop}%
	\bibitem [{\citenamefont {Matyjasek}\ and\ \citenamefont
		{Opala}(2017)}]{Matyjasek:2017psv}%
	\BibitemOpen
	\bibfield  {author} {\bibinfo {author} {\bibfnamefont {J.}~\bibnamefont
			{Matyjasek}}\ and\ \bibinfo {author} {\bibfnamefont {M.}~\bibnamefont
			{Opala}},\ }\href {\doibase 10.1103/PhysRevD.96.024011} {\bibfield  {journal}
		{\bibinfo  {journal} {Phys. Rev. D}\ }\textbf {\bibinfo {volume} {96}},\
		\bibinfo {pages} {024011} (\bibinfo {year} {2017})},\ \Eprint
	{http://arxiv.org/abs/1704.00361} {arXiv:1704.00361 [gr-qc]} \BibitemShut
	{NoStop}%
\end{thebibliography}
%

\end{document}